\documentclass[final]{ws-rv9x6}
\usepackage{subfigure} 
\usepackage{ws-rv-thm} 
\usepackage{hyperref}
\usepackage{url}
\usepackage[square]{ws-rv-van}
\makeindex

\definecolor{darkblue}{RGB}{46,48,147}
\hypersetup{colorlinks=true,
            linkcolor=blue,
            urlcolor=blue,
            citecolor=blue}

\DeclareMathOperator*{\arctanh}{arctanh}

\DeclareMathOperator*{\concat}{concat}
\DeclareMathOperator*{\softmax}{softmax}

\makeatletter\@addtoreset{chapter}{part}\makeatother%
\begin{document}

\chapter{Graph Neural Networks for Particle Tracking and Reconstruction}
\label{AIHEP:5.3}

\author{Javier Duarte}
\address{University of California San Diego, La Jolla, CA 92093}
\author[J. Duarte \& J-R. Vlimant]{Jean-Roch Vlimant}
\address{California Institute of Technology, Pasadena, CA 91125}

\begin{abstract}
Machine learning methods have a long history of applications in high energy physics (HEP).
Recently, there is a growing interest in exploiting these methods to reconstruct particle signatures from raw detector data.
In order to benefit from modern deep learning algorithms that were initially designed for computer vision or natural language processing tasks, it is common practice to transform HEP data into images or sequences.
Conversely, graph neural networks (GNNs), which operate on graph data composed of elements with a set of features and their pairwise connections, provide an alternative way of incorporating weight sharing, local connectivity, and specialized domain knowledge.
Particle physics data, such as the hits in a tracking detector, can generally be represented as graphs, making the use of GNNs natural.
In this chapter, we recapitulate the mathematical formalism of GNNs and highlight aspects to consider when designing these networks for HEP data, including graph construction, model architectures, learning objectives, and graph pooling.
We also review promising applications of GNNs for particle tracking and reconstruction in HEP and summarize the outlook for their deployment in current and future experiments.
\end{abstract}
\body

\tableofcontents

\section{Introduction}
\label{sec:intro}

Since the 1980s, machine learning (ML) techniques, including boosted decision trees, support vector machines, cellular automata, and multilayer perceptrons, have helped shape experimental particle physics~\cite{Denby:1987rk,Radovic:2018dip}.
As deep neural networks have achieved human-level performance for various tasks such as object recognition in images, they have been adopted in the physical sciences~\cite{Carleo:2019ptp} including particle physics.
Unlike traditional approaches, deep learning techniques operate on lower-level information to extract higher-level patterns directly from the data.
Applications of ML in high energy physics (HEP) have skyrocketed in recent years~\cite{Radovic:2018dip,Guest:2018yhq,Bourilkov:2019yoi,Larkoski:2017jix,livingreview}.
However, until recently it was necessary to completely transform HEP data into images or sequences in order to use modern deep learning algorithms that were initially designed for computer vision or natural language processing tasks.

Geometric deep learning (GDL)~\cite{4700287,Battaglia:2016jem,pointnet,bronstein2017geometric,gilmer2017neural,DGCNN,battaglia2018relational} is a growing subfield of artificial intelligence (AI) that studies techniques generalizing structured deep neural network models to non-Euclidean domains such as sets, graphs, and manifolds. 
This includes the study of graph neural networks (GNNs) that operate on graph data composed of elements with a set of features, and their pairwise connections.
Extensive reviews of GNNs are available in Ref.~\cite{bronstein2017geometric,battaglia2018relational,zhou2018graph,wu2019comprehensive,zhou2019graph,zhang2020deep,Bacciu_2020} that provide indepth technical details of current models.

As the data from particle physics experiments are generally sparse samplings of physics processes in time and space, they are not easily represented as regular-grid images or as ordered sequences.
Moreover, to reconstruct the input measurements into target particles, there is not always a clean, one-to-one mapping between the set of measurements and the set of particles because one particle can leave multiple traces in different subdetectors (many-to-one) and multiple particles can contribute to the same signal readout (one-to-many).
GDL algorithms, including GNNs, are well-suited for this type of data and event reconstruction tasks.
Unlike fully-connected (FC) models, convolutional neural networks (CNNs), and recurrent neural networks (RNNs), GNNs fully exploit the relational structure of the data. 
Recent work has applied set- and graph-based architectures in the domain of particle physics to charged particle tracking~\cite{Farrell:2018cjr,Ju:2020xty}, jet classification~\cite{neuraljets,Komiske:2018cqr,Moreno:2019neq,Moreno:2019bmu,Qu:2019gqs,Chakraborty:2020yfc,Bernreuther:2020vhm,Dolan:2020qkr} and building~\cite{Guo:2020vvt,Ju:2020tbo}, event classification~\cite{Abdughani:2018wrw,Ren:2019xhp,Abdughani:2020xfo}, clustering~\cite{Qasim:2019otl,Ju:2020xty}, vertexing~\cite{Serviansky:2020qwa,Shlomi:2020ufi}, particle finding~\cite{Kieseler:2020wcq}, and pileup mitigation~\cite{Martinez:2018fwc,Mikuni:2020wpr}.
Many of these applications are reviewed in Ref.~\cite{Shlomi:2020gdn}.

Analyses in particle physics are usually performed on high-level features, abstracted from the low-level detector signals.
The distillation of the raw detector data into a physics-centric representation is called \textit{reconstruction}, and is traditionally done in multiple stages---often at different levels of abstraction that physicists can naturally comprehend.
A classic reconstruction algorithm, by design, may be limited in how much detail and information is used from the data, often to simplify its commissioning and validation. 
Conversely, an algorithm based on ML can learn directly from the full complexity of the data and thus may potentially perform better. 
This effect is well illustrated in the sector of \textit{jet tagging}~\cite{AIHEP:6.3}
, where ML has brought significant improvements~\cite{Larkoski:2017jix}.
GNNs, because of the relational inductive bias they carry, have a great deal of expressive power when it comes to processing graph-like objects.
However, there is a delicate balance between the increased expressivity and the incurred computational cost.

\begin{figure}[htpb]
    \centering
    \subfigure[]{
    \includegraphics[width=\textwidth]{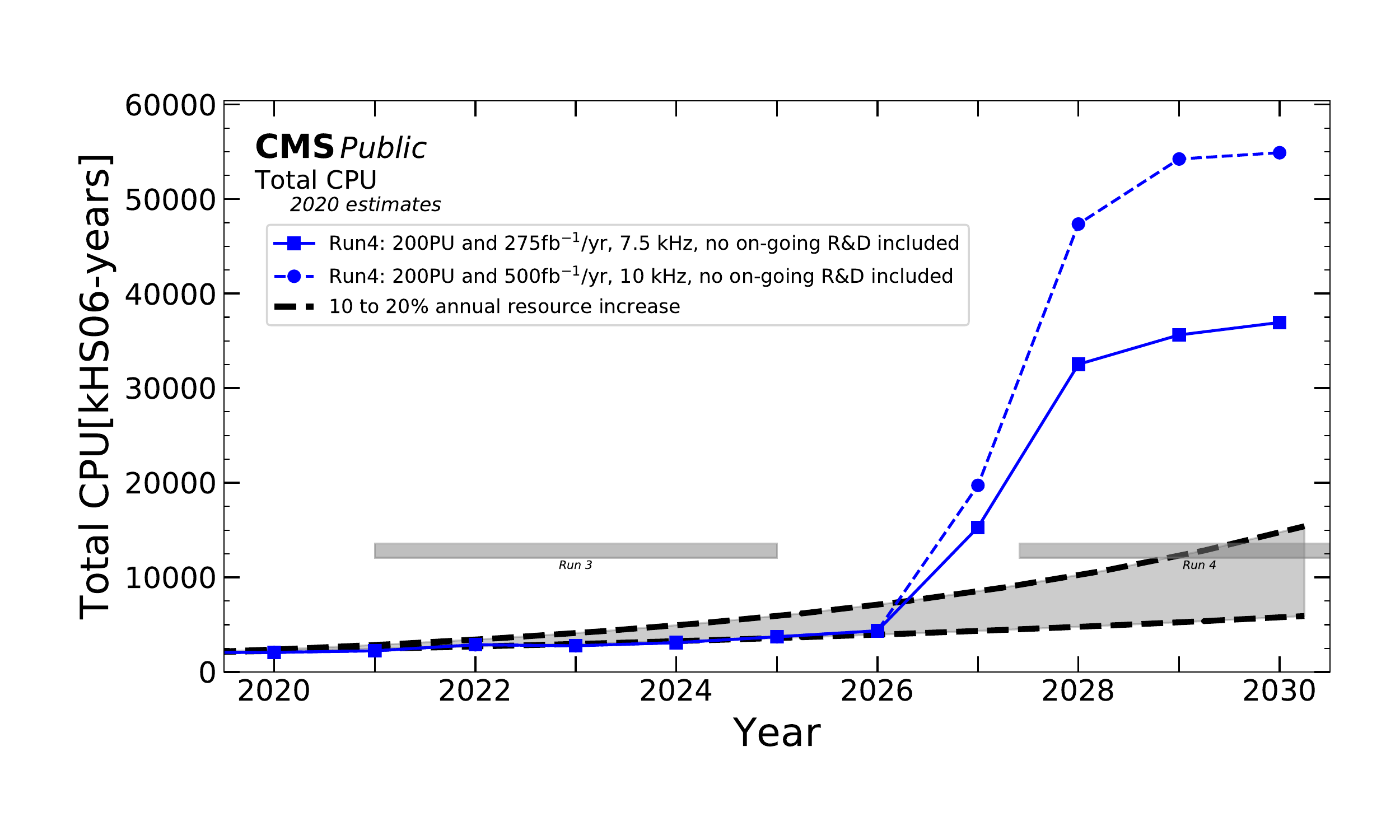}}
    \subfigure[]{
    \includegraphics[width=0.8\textwidth]{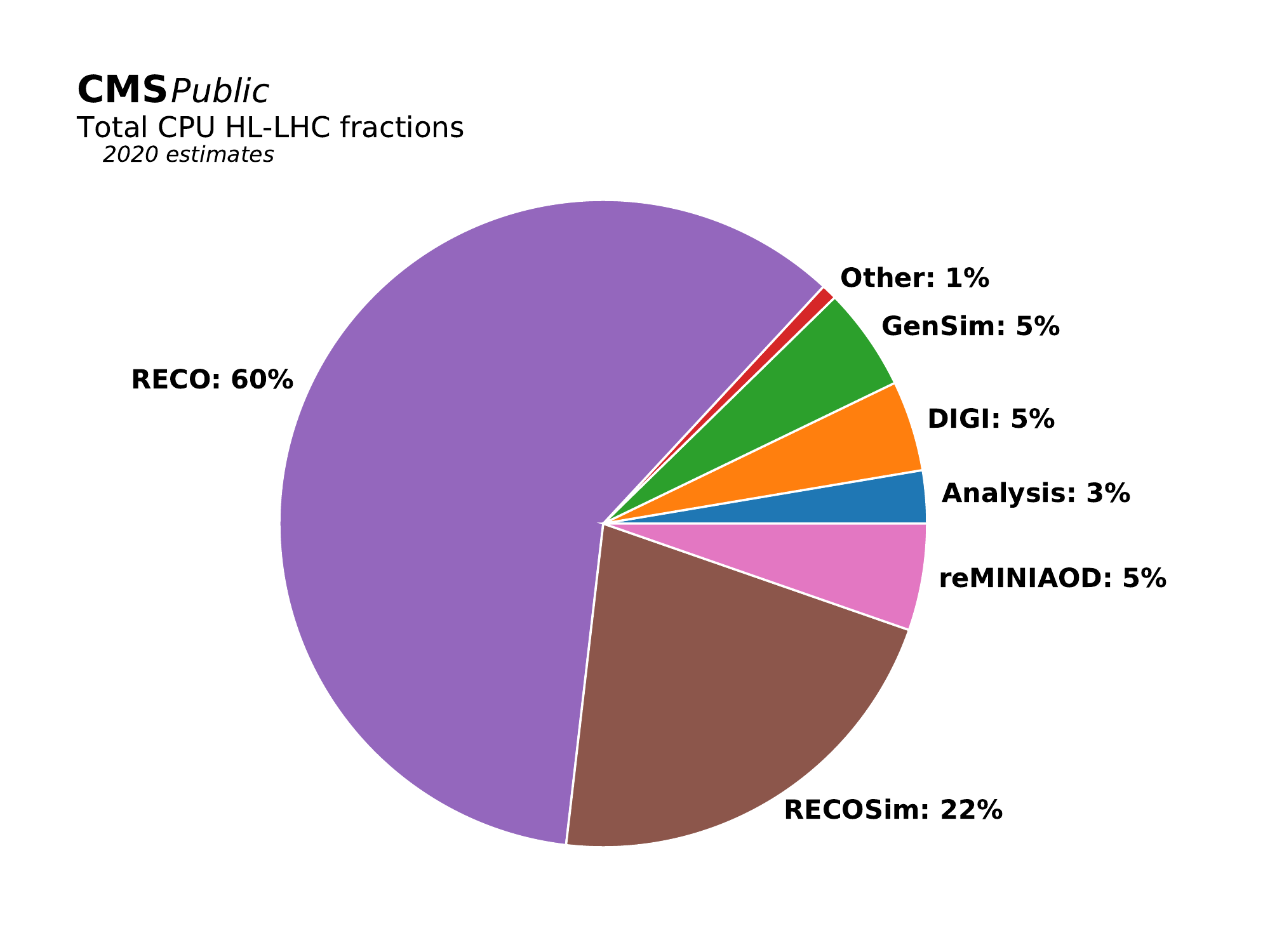}}
    \caption{CPU time annual requirements (in kHEPS06-years) estimated for CMS processing and analysis needs (a)~\cite{Alves:2017she,octwiki}.
    kHS06-years stands for $10^3$~HEPSPEC06 per year, a standard CPU performance metric for HEP.
    Two scenarios are considered: one that assumes reaching 275~fb$^{-1}$ per year during Run 4 with 7.5~kHz of data saved and a second that assumes reaching 500~fb$^{-1}$ per year during Run 4 with 10~kHz of data saved (dashled line). 
    The blue curves (and points) show the annual projected CPU need, summed across Tier-0, Tier-1 and Tier-2 resource needs in each of these scenarios. 
    The black curve shows the projected resource availability extrapolating the current CMS processing resources assuming an annual increase of 10--20\%. 
    Approximate breakdown of CPU time requirements into primary processing and analysis activities for the first scenario (b)~\cite{Alves:2017she,octwiki}.
    \label{fig:comphllhc}}
\end{figure}

A significant motivation for studying novel ML algorithms for reconstruction, especially charged particle tracking, is their large computational burden for big data HEP experiments.
Figure~\ref{fig:comphllhc} shows the large increase of expected computational resources needed for all activities in the CMS experiment after the planned major upgrade of the LHC.
The largest fraction (60\%) of CPU time is consumed by reconstruction-related tasks and of this, the largest component belongs to tracking.
The complexity of the current reconstruction algorithms with respect to increasing event density is such that we foresee future shortcomings in computing resources.
Several factors contribute to the slowdown in the evolution of single-core CPU performance~\cite{dennard,breakdown}, and highly parallel architectures like graphics processing units (GPUs) now provide more of the computing power in modern high-performance computing centers.
While some reconstruction algorithms already take advantage of multithreaded optimizations~\cite{Jones:2011za,Hernandez:2012up,Jones:2015soc,Riley:2019ebh}, it is a major endeavor to fully migrate the software to highly parallel architectures~\cite{Bocci:2020olh}.
Deep learning models offer a natural way to take advantage of GPUs in production.
By leveraging greater parallelism, an ML-based algorithm might execute faster with a smaller computational footprint than a traditional counterpart even though it may require more floating point operations (FLOPs).
In this way, the complexity of ML-based algorithms---including the pre-processing and post-processing steps---may be better than that of existing counterparts.


This chapter is structured as follows. 
Sec.~\ref{sec:graphdata} provides an overview of the different ways that particle physics data may be encoded as graphs.
In Sec.~\ref{sec:gnns}, we recapitulate the formalism behind commonly used GNNs.
In Sec.~\ref{sec:discuss}, we highlight several design considerations, including computational performance, for various approaches to building GNNs for HEP reconstruction.
In~\ref{sec:gnnapps}, we review the suite of GNN applications to tracking and reconstruction tasks.
Finally, we summarize the chapter in Sec.~\ref{sec:summary}.

\section{Point Cloud and Graph Data}

\label{sec:graphdata}

Modern detectors are an assembly of several different technologies with a wide range of spatial granularities (down to $\mathcal O(1)$~mm) and a total size of $\mathcal O(10)$~m.
Therefore, the signals from the detector are extremely heterogeneous.
In many cases, the measurements are inherently sparse because of the event configurations of the physics processes.
At the same time, the local density of the measurements can be extremely high because of the fine granularity of the active material, for example in the tracker.
The signal is also sampled in time, although for most detectors, it is effectively discretized in units of one beam crossing period, which is 25~ns for the LHC.

Locally, a fraction of the data, especially from the calorimeters, can be interpreted as images.
In particular, jet images~\cite{Cogan:2014oua} are a now-common representation of localized hadron showers in calorimeters. 
This has led to proliferation of image-based deep learning techniques, such as CNNs, skip connections, or capsules, for calorimeter- or jet-related tasks with substantial performance improvements over traditional methods~\cite{ATLAS:2017dfg,Kasieczka:2017nvn,Macaluso:2018tck,Andrews:2018nwy,Lin:2018cin,ATLAS:2019fxb}.
However, the image-based representations face some stringent limitations due to the irregular geometry of detectors and the sparsity of the input data.
Alternatively, a subset of detector measurements and reconstructed objects can be interpreted as ordered sequences. 
Methods developed for natural language processing, including RNNs, long-short term memory (LSTM) cells, or gated recurrent units (GRUs), may therefore be applied~\cite{ATLAS:2017gpy,Sirunyan:2020lcu}.
While the ordering can usually be justified experimentally or learned~\cite{Louppe:2017ipp}, it is often arbitrary and constrains how the data is presented to models.

\begin{figure}[htbp]
    \centering
    \includegraphics[height=0.4\textwidth]{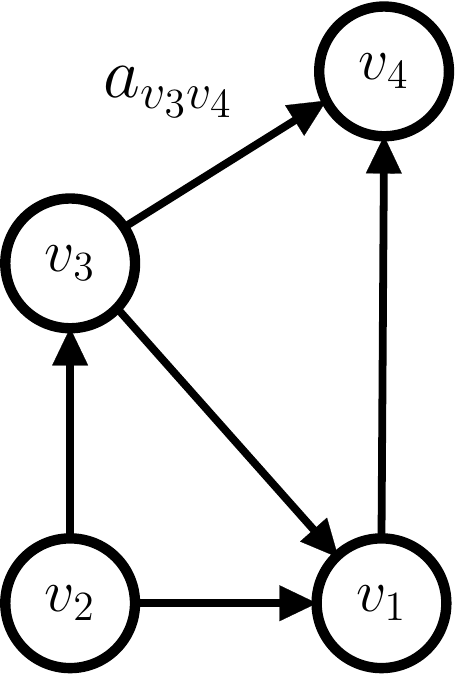}
    \includegraphics[height=0.4\textwidth]{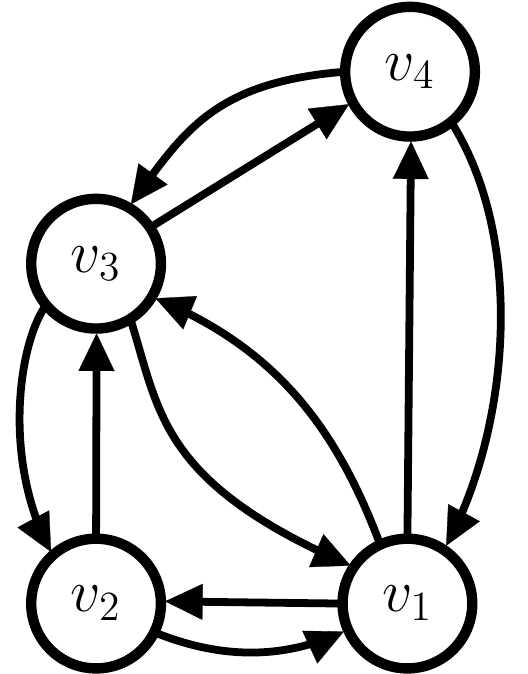}
    \caption{A directed graph with oriented arcs (left). 
    If the graph is undirected, it can be transformed into a directed one to obtain a viable input for graph learning methods (right).
    In particular, each edge is replaced by two oriented and opposite arcs with identical edge features~\cite{Bacciu_2020}.}
    \label{fig:directed}
\end{figure}

Fundamentally, the raw data is an unordered set of $N^v$ items. 
However, by additionally considering $N^e$ geometric or physical relationships between items (encoded by an \emph{adjacency matrix}), the set can be augmented into a graph.
These relationships may be considered directed or undirected as shown in Fig~\ref{fig:directed}.
An adjacency matrix is a (typically sparse) binary $N^v\times N^v$ matrix, whose elements indicate whether a given vertex is adjacent to another vertex.
Another, equivalent representation is through an $N^v\times N^e$ \emph{incidence} matrix, whose elements indicate whether a given vertex is connected to a given edge.
A third alternative encoding of an adjacency matrix is in coordinate list (COO) format, i.e. a $2\times N^e$ matrix where each column contains the node indices of each edge.
This compact representation is beneficial in terms of incremental matrix construction and reduced size in memory, but for arithmetic operations or slicing a conversion to a compressed sparse row (CSR), compressed sparse column (CSC), or dense format is often necessary.

\begin{figure}[htbp]
	\centering
	\subfigure[]{\includegraphics[width=\textwidth]{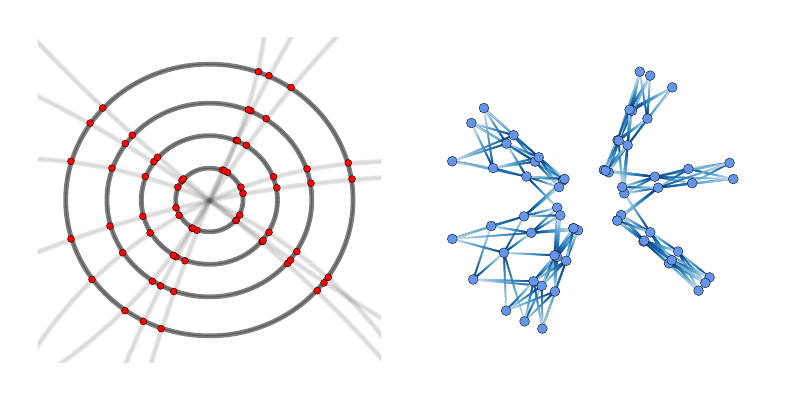}}\\
	\subfigure[]{\includegraphics[width=\textwidth]{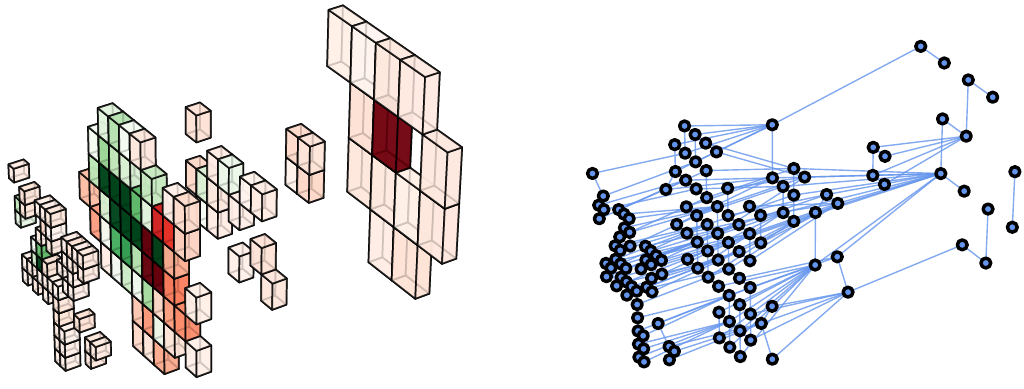}}
    \caption{HEP data lend themselves to graph representations for many applications: segments of hits in a tracking detector hits (a), and neighboring energy deposits in calorimeter cells (b). 
    Figures reproduced from Ref.~\cite{Shlomi:2020gdn}.}
		\label{fig:data_as_graph}
\end{figure}

A graph representation is more flexible and general than images or sequences. 
In particular, one may recover an image or sequence representation by appropriate choice of the adjacency matrix.
Moreover, there may be less preprocessing required to apply deep learning to this representation of the data.
For example, for an image representation of calorimeter hit data, it may be necessary to first cluster the hits, form the two-dimensional energy-weighted image, and center, normalize, rescale, or rotate the image~\cite{Cogan:2014oua,deOliveira:2015xxd}.
These manipulations of the data may have undesirable consequences, including loss of particle-level information, distortions of physically meaningful information like jet substructure, modifying Lorentz-invariant properties of the data (e.g. particle mass), and imposing translational invariance in $\eta$-$\phi$ space, which does not respect this symmetry~\cite{Pearkes:2017hku}.
In contrast, a GNN, may be able to operate on the unclustered hit data, with appropriately chosen connections, directly.
Two example HEP detector data sets and their possible graph encoding are illustrated in Fig.~\ref{fig:data_as_graph}.

\subsection{Graph Construction}
\label{sec:graphconstruction}

\begin{figure}[htpb]
    \centering
    \includegraphics[width=\textwidth]{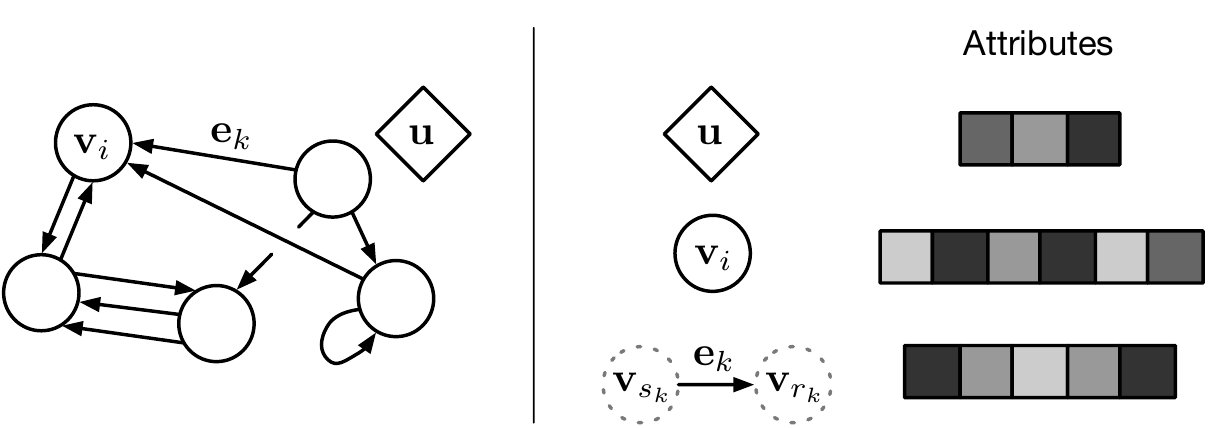}
    \caption{A directed, attributed multi-graph $\mathcal{G}$ with a global attribute~\cite{battaglia2018relational}. 
    A node is denoted as $\mathbf{v}_i$, an edge as $\mathbf{e}_k$, and the global attributes as $\mathbf{u}$. 
    The indices $s_k$ and $r_k$ correspond the sender and receiver nodes, respectively, for the one-way edge $k$ (from the sender node to the receiver node).}
    \label{fig:attributes}
\end{figure}

In particle physics applications, the specific relationships between set elements to present to an algorithm depends on the context and objective. 
Subjective choices must be made to construct a graph from the set of inputs. 
Formally, a graph is represented by a triplet $\mathcal G = (\mathbf{u}, V, E)$, consisting of a graph-level, or \textit{global}, feature vector $\mathbf{u}$, a set of $N^v$ nodes $V$, and a set of $N^e$ edges $E$.
The nodes are given by $V = \{\mathbf{v}_i\}_{i=1:N^v}$, where $\mathbf{v}_i$ represents the $i$th node's attributes.
The edges connect pairs of nodes, $E = \{\left(\mathbf{e}_k, s_k, r_k\right)\}_{k=1:N^e}$, where $\mathbf{e}_k$ represents the $k$th edge's attributes, and $s_k$ and $r_k$ are the vectors of indices of the ``sender'' and ``receiver'' nodes, respectively, connected by the $k$th edge (from the sender to the receiver node).
The receiver and sender index vectors are an alternative way of encoding the directed adjacency matrix, as discussed above.
The graph and its attributes are represented pictorially in Fig.~\ref{fig:attributes}.
Edges in the graph serve three different functions:
\begin{enumerate}
	\item the edges are communication channels among the nodes,
	\item input edge features can encode a relationship between objects, and
	\item latent edges store relational information learned by the GNN that are relevant for the task.
\end{enumerate} 
Depending on the task, creating pairwise relationships between nodes may even be entirely avoided, as in the deep sets~\cite{deepsets,Komiske:2018cqr} architecture with only node and global properties.

\begin{figure}[htbp]
	\centering
	\includegraphics[width=0.8\textwidth]{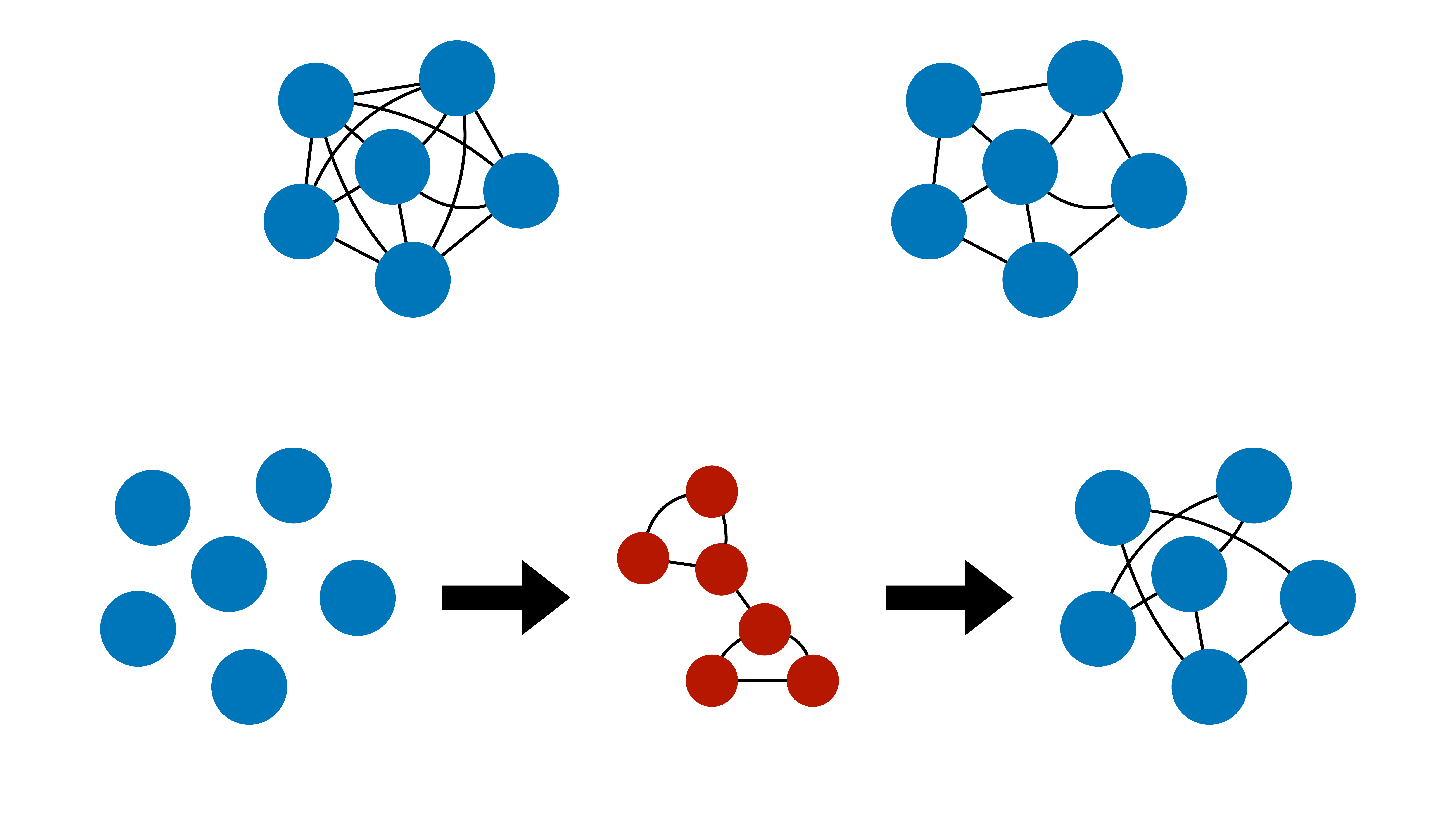}
	\caption{Different methods for constructing the graph: connecting all pairs of nodes (upper left), connecting neighboring nodes in a predefined feature space (upper right), and connecting neighboring nodes in a latent feature space (lower).
	\label{fig:graphconstruction}}
\end{figure}

For small input sets, with $N^v<100$, a simple choice is to form a fully-connected graph, allowing the network to learn about all possible object relationships.
As the number of edges in a fully-connected graph increases as $N^e \propto ({N^v})^2$, the computational cost of applying a neural network to all of the edges becomes prohibitive. 
A work-around is to precompute a fixed edge feature, such as the geometric distance between nodes, that can be focus on certain neighboring nodes. 

If edge-level computations is required, it may be necessary to restrict the considered edges.
Edges can be formed based on the input features (e.g. the $\Delta R = \sqrt{\Delta \phi^2 + \Delta \eta^2}$ between particles) or a learned representation, such as that used by the EdgeConv~\cite{DGCNN,Qu:2019gqs} and GravNet~\cite{Qasim:2019otl} architectures.
Given a distance metric between nodes and a criterion for connecting them, such as $k$-nearest neighbors or a fixed maximum distance, the edges can be created.
These three different graph construction methods are illustrated in Fig.~\ref{fig:graphconstruction}.


\section{Graph Neural Networks}
\label{sec:gnns}

GNNs are a class of models for reasoning about explicitly structured data, in particular
graphs~\cite{4700287,santoro2017simple,gilmer2017neural,bronstein2017geometric,wang2018nonlocal,li2018learning,kipf2018neural}.
These approaches all share a capacity for performing computation over discrete entities and the relations between them.
Crucially, these methods carry strong relational inductive biases, in the form of specific architectural assumptions, which guide these approaches towards learning about entities and relations~\cite{Mitchell80}.

Here, we recapitulate the ``graph network'' (GN) formalism~\cite{battaglia2018relational}, which synthesizes various GNN methods.
Fundamentally, GNs are graph-to-graph mappings, whose output graphs have the same structure as the input graphs. 
Formally, a GN block contains three ``update'' functions, $\phi$, and three ``aggregation'' functions, $\rho$.
The stages of processing in a single GN block are:
\begin{align}
&\text{(Aggregation)} & &\text{(Update)} \nonumber \\
    &&\mathbf{e}'_k &= \phi^e\left(\mathbf{e}_k, \mathbf{v}_{r_k}, \mathbf{v}_{s_k}, \mathbf{u} \right) & \text{(Edge block),}\\
    \mathbf{\bar{e}}'_i &= \rho^{e \rightarrow v}\left(E'_i\right) & \mathbf{v}'_i &= \phi^v\left(\mathbf{\bar{e}}'_i, \mathbf{v}_i, \mathbf{u}\right) &  \text{(Node block),}\\
    \begin{split}
        \mathbf{\bar{e}}' &= \rho^{e \rightarrow u}\left(E'\right) \\
        \mathbf{\bar{v}}' &= \rho^{v \rightarrow u}\left(V'\right) 
    \end{split} &
    \mathbf{u}' &= \phi^u\left(\mathbf{\bar{e}}', \mathbf{\bar{v}}', \mathbf{u}\right) &\text{(Global block).}
  \label{eq:gn-functions}
\end{align}
where $E'_i = \left\{\left(\mathbf{e}'_k, r_k, s_k \right)\right\}_{r_k=i,\; k=1:N^e}$ contains the updated edge features for edges whose receiver node is the $i$th node, $E' = \bigcup_i E_i' = \left\{\left(\mathbf{e}'_k, r_k, s_k \right)\right\}_{k=1:N^e}$ is the set of updated edges, and $V'=\left\{\mathbf{v}'_i\right\}_{i=1:N^v}$ is the set of updated nodes.
We describe each block below.

\begin{figure}
    \centering
    \includegraphics[width=\textwidth]{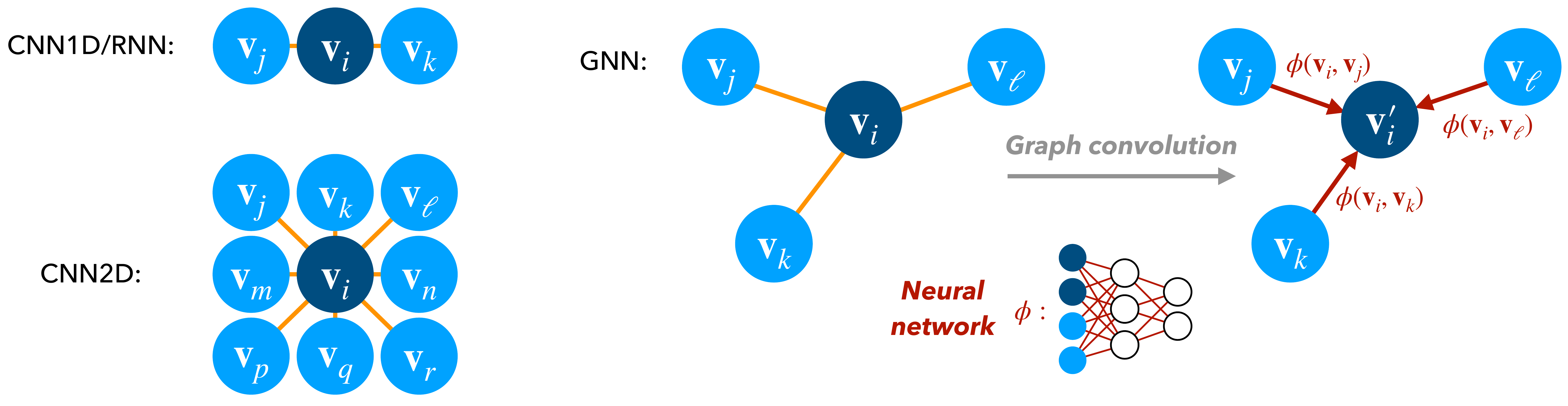}
    \caption{Input structure for 1D CNNs and RNNs (top left) and 2D CNNs (bottom left) compared to GNNs (right). 
    In a 2D convolution, each pixel in an image
is taken as a node where a fixed number of neighbors are determined by their proximity and  filter size. 
    RNNs compute sequentially along the input data, generating a sequence of hidden states, as a function of the previous hidden state and the input.
    A graph convolution operation applies a pair-wise neural network to a variable-size and unordered set of neighboring nodes, and then aggregates the results.}
    \label{fig:edgeconv}
\end{figure}

The \textit{edge block} computes an output for each edge $\mathbf{e}'_k$, known as the updated edge feature or ``message.''
These are subsequently aggregated according to the corresponding receiver nodes $\mathbf{\bar{e}}'_i = \rho^{e \rightarrow v}\left(E'_i\right)$ in the first part of the \textit{node block}. 
These two steps are sometimes known as the graph or edge convolution or message-passing operation. 
In some ways, this operation generalizes the type of convolution done in CNNs, and the sequential, recurrent processing of RNNs, as shown in Fig.~\ref{fig:edgeconv}.
In a 2D convolution, each pixel in an image is processed together with a fixed number of neighboring pixels determined by their spatial proximity and the filter size. 
RNNs compute sequentially along the input data, generating a sequence of hidden states $\vec{h}_t$, as a function of the previous hidden state $\vec{h}_{t-1}$ and the input for position $t$.
In contrast, a graph convolution operation applies a pair-wise neural network to all neighboring nodes, and then aggregates the results to compute a new hidden representation for each node $\mathbf{v}'_i$.
As opposed to image and sequence data, the neighbors of a node in a graph are unordered and variable in number.

As described above, the aggregation function $\rho^{e\to v}$ maps edge-specific information to node-specific outputs by compiling information based on the receiver node indices.
To apply generically to unordered graph-structured data, the $\rho$ functions must be invariant to permutations of their inputs, and should take variable numbers of arguments.
Examples include an element-wise summation, mean, maximum, and minimum.
This construction ensures permutation invariance of the GNN as a whole.
In Ref.~\cite{wagstaff2019limitations}, it was shown that this invariance suggests a minimum size for the latent dimension: for scalar inputs the dimensionality of $\phi$ has to be at least equal to the number of inputs (i.e. nodes or edges) in order to be able to approximate any permutation-invariant function.
Other authors have also considered permutation- and group-equivariant constructions~\cite{kondor2018generalization,maron2019invariant,keriven2019universal,sannai2019universal,Bogatskiy:2020tje,miller2020relevance,Smidt2020,dym2020universality}, which are not covered here.

The rest of the node block computes an output for each node $\mathbf{v}'_i = \phi^v\left(\mathbf{\bar{e}}'_i, \mathbf{v}_i, \mathbf{u}\right)$. 
This can be thought of as an update of the node features, which takes into account the previous node features, the global features, and one round of message passing among neighboring nodes.
That is, relational information from nearest neighbors in the graph are used to update the node features.

\begin{figure}[htbp!]
\centering
	\includegraphics[width=\textwidth]{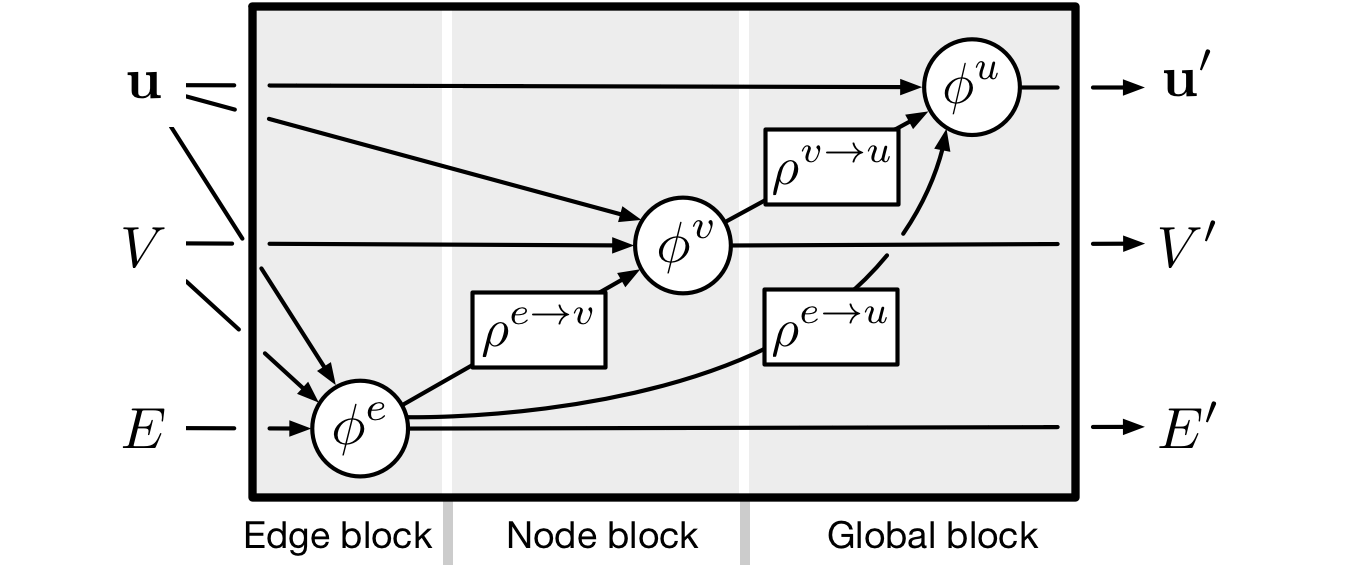}
	\caption{A GN block from Ref.~\cite{battaglia2018relational} that processes an input graph $G=(\mathbf{u}, V, E)$ and returns a graph with the same structure but updated attributes $G'=(\mathbf{u}', V', E')$.
	\label{fig:gn-framework}}
\end{figure}

Finally, the edge- and node-level outputs are each aggregated with $\rho^{e\to u}$ and $\rho^{v\to u}$, respectively, in order to compute graph-level information in the \textit{global block}. 
The output of the GN is the triplet of updated edge, node, and global features, $\mathcal G'=(\mathbf{u}', V', E')$ as shown in Fig.~\ref{fig:gn-framework}.

The GN formalism is generic for graph-to-graph mappings.
GNs also generalize to graphs not seen during training, because the learning is focused at the edge- and node-level, although such generalization may require conditions to be satisfied between the training and test graph domains~\cite{yehudai2020size,NEURIPS2019_4c5bcfec,10.1145/3292500.3330956}.
Except for the global block, the GN never considers the full graph in a computation.
Nonetheless, when multiple GN blocks are stacked in deep or recurrent configurations, information can propagate across the graph's structure, allowing more complex, long-range relationships to be learned.

As an example of the generality of the GN framework, it can be used to express the dynamic edge convolution (EdgeConv) operation of the dynamic graph CNN (DGCNN)~\cite{DGCNN}, which is commonly used in HEP.
This layer operates on a graph selected using the $k$-nearest neighbors of the nodes, including self-loops.
Edge features are computed as 
\begin{equation}
\mathbf{e}'_k = \phi^e(\mathbf{v}_{r_k}, \mathbf{v}_{r_k} - \mathbf{v}_{s_k}).
\end{equation}
The choice of $\phi^e$ adopted in Ref.~\cite{DGCNN} is an asymmetric edge function that explicitly combines the global shape structure, captured by the coordinates $\mathbf{v}_{r_k}$, with local neighborhood information, captured by $\mathbf{v}_{r_k} - \mathbf{v}_{s_k}$.
The EdgeConv operation also uses a permutation-invariant aggregation operation $\rho^{e\to v}$ (e.g., $\sum$ or $\max$) on the edge features associated with all the edges emanating from each node. 
The output of the EdgeConv operation at the $i$th node is thus given by %
\begin{equation}
\label{eq:full}
\mathbf{v}'_i  = \phi^v(\mathbf{\bar e}'_i) = \mathbf{\bar e}'_i ,
\end{equation}
that is the $\phi^v$ function is trivial.
A crucial difference with the GN framework is that after each EdgeConv layer, the connectivity of the graph is recomputed using the $k$-nearest neighbors in the latent space. 
This dynamic graph update is the reason for the name of the architecture.
Similarly, GravNet and GarNet~\cite{Qasim:2019otl} are two  other GNN architectures that use the distance in a latent space when aggregating to predict a new set of node features. 

Other GNN models are also expressible within this framework or with minor modifications.
For instance, interaction networks~\cite{Battaglia:2016jem} use a full GN block except for the absence of the global features to update the edge properties.
Deep sets~\cite{deepsets} bypass the edge update completely and predict the global output from pooled node information directly.
PointNet~\cite{pointnet} use similar update rule, with a max-aggregation for $\rho^{v\to u}$ and a two-step node update.

\begin{figure}
    \centering
  \includegraphics[scale=0.6]{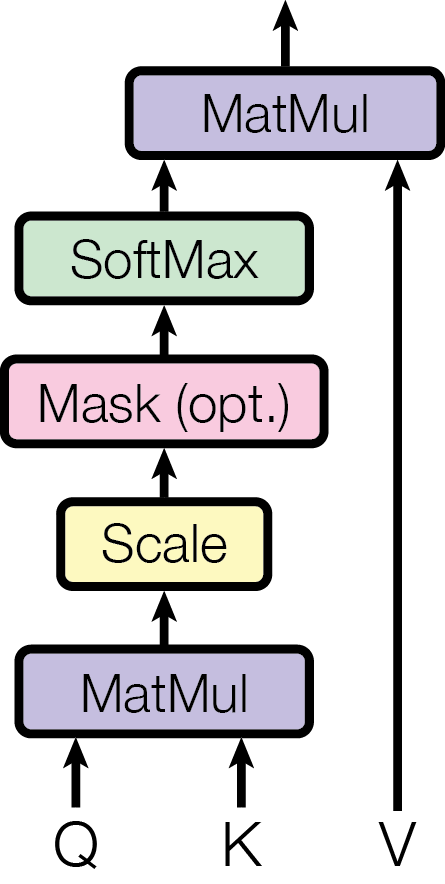}\hspace{0.15\textwidth}
  \includegraphics[scale=0.6]{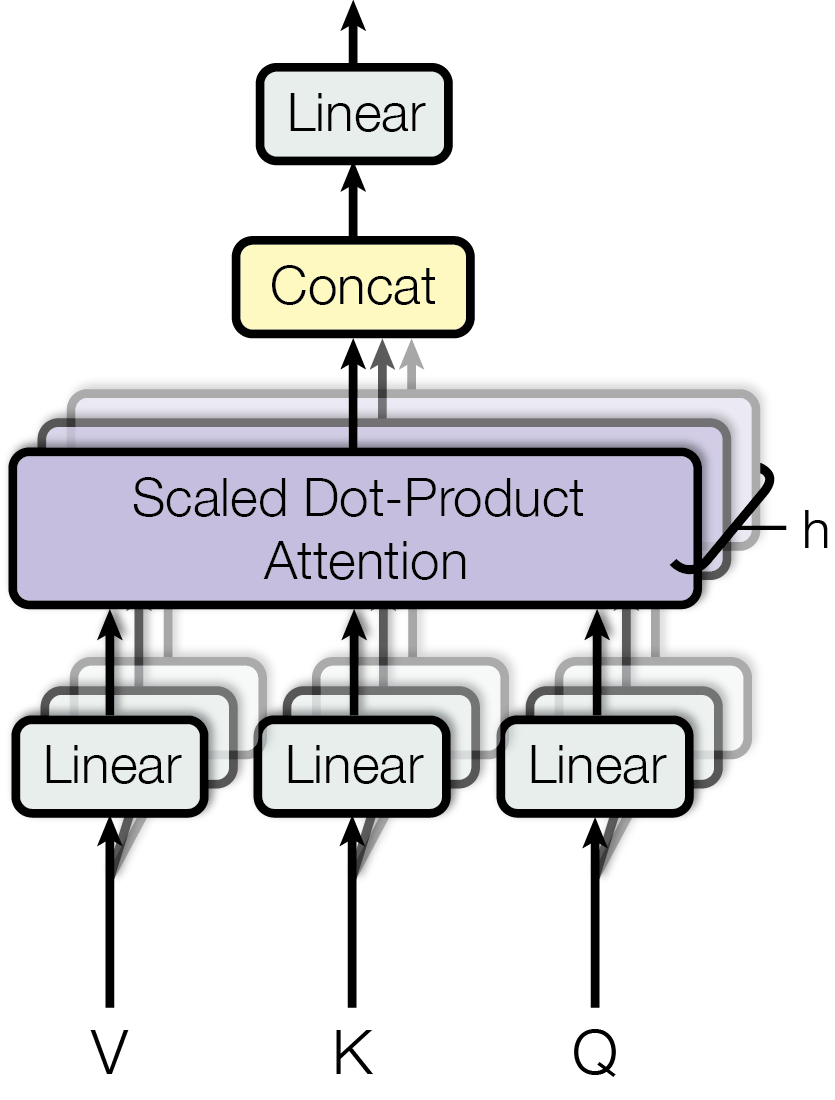}  
    \caption{Scaled dot-product attention (left) and multi-head attention (right), consisting of several attention layers running in parallel, from Ref.~\cite{vaswani2017attention}.}
    \label{fig:attention}
\end{figure}

Another class of models closely related to GNNs that perform predictions on structured data, especially sequences, are \emph{transformers}, based on the self-attention mechanism~\cite{vaswani2017attention}.
At a high level, a self-attention layer is a mapping from an input sequence, represented as a $n\times d_\mathrm{in}$ matrix $X$ (where $n$ is the sequence length and $d_\mathrm{in}$ is the dimensionality of the input features) to a $n\times d_\mathrm{out}$ output matrix through an attention function, which focuses on certain positions of the input sequence.
A self-attention function takes as input an $n\times d_k$ query matrix $Q$, and a set of key-value pairs, represented by a $n\times d_k$ matrix $K$ and a $n\times d_v$ matrix $V$, respectively, all of which are transformed versions of the input sequence
\begin{equation}
Q=XW_Q, K=XW_K, V=XW_V,
\end{equation}
where $W_Q$, $W_K$, and $W_V$ are learnable $d_\mathrm{in}\times d_k$, $d_\mathrm{in}\times d_k$, and $d_\mathrm{in}\times d_\mathrm{out}$ matrices, respectively.
The scaled dot-product attention (see Fig.~\ref{fig:attention}) is computed by taking the dot products of the query with all keys (as a compatibility test) divided by $\sqrt{d_k}$ and applying a softmax function to obtain the weights for the values. 
In matrix form:
\begin{align}
   \mathrm{Attention}(Q, K, V) &= \softmax\left(\frac{QK^\top}{\sqrt{d_k}}\right)V.
\end{align}
An important variant of this is \emph{multi-head attention} depicted in Fig.~\ref{fig:attention}: instead of applying a single attention function, it is beneficial to project the queries, keys, and values $h$ times into subspaces whose dimensions are $h$ times smaller.
On each of these projected versions of queries, keys, and values, the attention function is computed yielding $h$ $d_v$-dimensional output values.
These are concatenated and once again projected, resulting in the final values: 
\begin{align}
    \mathrm{MultiHead}(X) &= \concat_{i\in [h]}[H^{(i)}]W^O\\
    \mathrm{where}~H^{(i)} &= \mathrm{Attention}(XW_Q^{(i)}, XW_K^{(i)}, XW_V^{(i)}),
\end{align}
and $W^O$ is a learnable $hd_v\times d_\mathrm{out}$ matrix.
In practice, a simplifying choice of $d_\mathrm{in} = hd_k = hd_v = d_\mathrm{out}$ is typically made.
Multi-head attention allows the model to jointly attend to information from different representation subspaces at different positions.

In the language of GNNs, a transformer computes normalized edge weights in a fully-connected graph, and passes messages along the edges that are aggregated in proportion to these weights. 
For example, the transformer in the graph attention network~\cite{gat} uses a $\phi^e$ function that produces both a vector message and an unnormalized weight.
The aggregator $\rho^{e\rightarrow v}$ then normalizes the weights before computing a weighted sum of the message vectors. 
This allows the edge structure among the input nodes to be inferred and used for message passing. 
In addition, attention mechanisms are a way to apply different weights in the aggregation operations $\rho$.

\begin{figure}
    \centering
    \includegraphics[width=\textwidth]{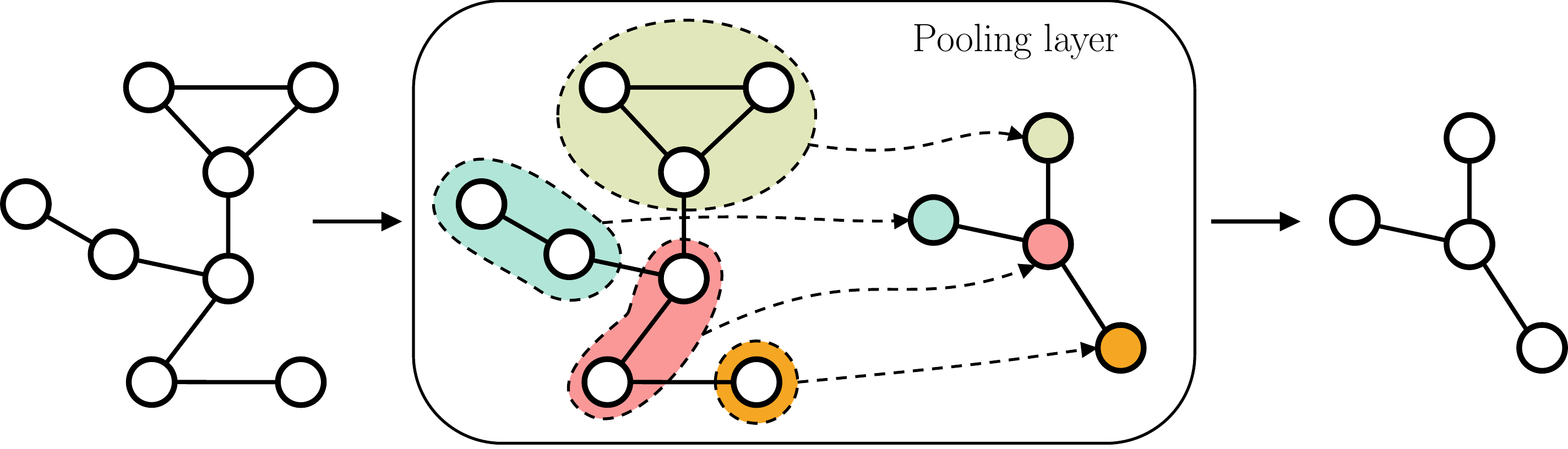}
    \caption{An example of a graph pooling layer that coarsens the graph by identifying and clustering nodes of the same neighborhood together, so that each group becomes a node of the coarsened graph~\cite{Bacciu_2020}.}
    \label{fig:pooling}
\end{figure}

Another extension of GNNs involves graph pooling, represented in Fig.~\ref{fig:pooling}.
Graph pooling layers play the role of ``downsampling,'' which coarsens a graph into a sub-structure.
Graph pooling is mainly used for three purposes: to discover important communities in the graph, to imbue this knowledge in the learned representations, and to reduce the computational costs of message passing in large scale structures.
Pooling mechanisms fall in two broad classes: adaptive and topological. 

Adaptive graph pooling relies on a parametric, trainable pooling mechanism. 
One example of this approach is differentiable pooling~\cite{ying2019hierarchical}, which uses a neural network layer to learn a clustering of the current nodes based on their embeddings at the previous layer.
Top-$k$ pooling~\cite{gao2019graph} learns node scores and retain only the entries corresponding to the top nodes. 
Node selection is made differentiable by means of a gating mechanism built on the projection scores. 
Self-attention graph (SAG) pooling~\cite{lee2019selfattention} extends top-$k$ pooling by using a GNN to learn attention scores.
Another example is edge pooling~\cite{edgepool}, in which edge scores are computed and edges are contracted iteratively according to those scores.
In contrast to these adaptive methods, topological pooling mechanisms are not required to be differentiable and typically leverage the structure of the graph itself.
The graph clustering software (GRACLUS)~\cite{graclus} implements a widely-used, efficient greedy clustering algorithm that matches vertices based on their edge weights.
Similarly, nonnegative matrix factorization pooling~\cite{bacciu2019nonnegative} provides a soft node clustering using a nonnegative factorization of the adjacency matrix.

\section{GNN Design Considerations}
\label{sec:discuss}

The formalism and methods introduced in Sec.~\ref{sec:gnns} expose the numerous dimensions of the space of GNN model architectures.
While the possibilities for combining the ingredients of GNN are limitless, other considerations and constraints come into play to shape the model for a given task and environment.
In this section, we discuss some of the salient facets of GNN design for HEP reconstruction tasks.
These are some of the guiding principles that lead to the models used for the applications we describe further in Sec.~\ref{sec:gnnapps}.

\subsection{Model Architectures}
\label{sec:models}


Many of the choices in the design the GNN model architectures reflect the learning objectives or aspects of the data that are specific to HEP.
The choice of architecture is an important way to incorporate inductive bias into the learning task.
For instance, this choice includes the size of the networks, the number of stacked GNN blocks, attention mechanisms, and different types of pooling or aggregation.
The model architecture should reflect a logical combination of the inputs towards the learning task. 
In the GN formalism, this means a concrete implementation of the block update and aggregation functions and their sequence.
As an example of such a choice, global aggregation can occur before a node update, or an edge representation can be created and aggregated to form a node update. 
The difference between the two is that one is based on a sum of pairwise representations, and the other on a global sum of node representations.

\begin{figure}[htpb]
	\centering
	\includegraphics[width=0.8\textwidth]{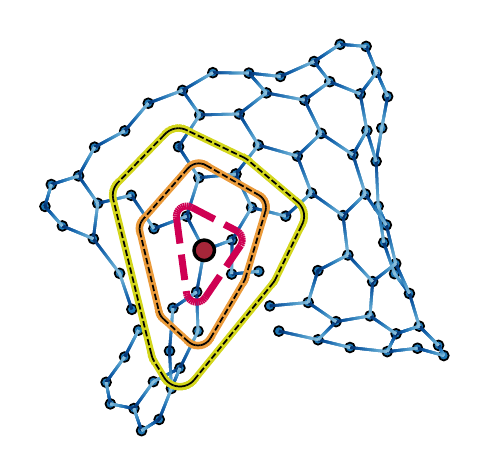}
	\caption{The red, orange-highlighted, and yellow-highlighted dotted lines represent the enlarging neighborhood of nodes that may communicate with the red node after one, two, and three iterations of message passing, respectively~\cite{Shlomi:2020gdn}.
	Those nodes outside of the yellow-highlighted dotted boundary do not influence the red node after three iterations.}
	\label{fig:message_passing}
\end{figure}

Stacks of GN blocks are also useful for two purposes. 
First, just as in CNNs, they can construct a higher-level, more abstract representation of the data.
Second, the number of iterations of message passing defines the nodes that can exchange information.
This is illustrated in Fig.~\ref{fig:message_passing}. 
Multiple iterations increase each nodes' neighborhood of communication, as the representation of its neighboring nodes was previously updated with messages from their neighbors.

Attention mechanisms also play an important role in emphasizing or deemphasizing certain nodes or connections during aggregation.
A popular choice is to use the $\Delta R$ distance between measurement nodes in the input space or Euclidean distance in the latent space (or subspace) as an edge weight. 
Others networks~\cite{Farrell:2018cjr} use the network's predicted edge weight, which acts to reinforce its learned connections.
Finally, the choice of aggregation method is crucial to keep open the appropriate communication channels and maintain the desired properties of the output, such as permutation invariance.

\subsection{Graph Reduction and Alternative Loss Functions}
\label{sec:losses}

One difficulty of applying deep learning to HEP data is the ``jagged'' or event-dependent nature of the target.
In particular, the number of physics objects, such as tracks, clusters, or final-state particles, to be reconstructed per event is variable and unknown apriori.
For this reason, methods based on a fixed output size for the output are challenging to apply.

Two methods~\cite{Kieseler:2020wcq,Gray:2020mcm} aim to specifically address this problem.
In Ref.~\cite{Kieseler:2020wcq}, a clustering or ``condensation'' of the input nodes is derived through a choice of condensation points and a dual prediction of a regression target and a condensation weight. 
The loss function is inspired by attractive and repulsive electromagnetic potentials, ensuring that nodes that belong to the same target object are kept close in the latent space.
Similarly, a dynamic reduction network is proposed in Ref.~\cite{Gray:2020mcm} uses a DGCNN~\cite{DGCNN} and a greedy popularity-based clustering algorithm~\cite{greedy} to reduce the number of nodes. 
The model was developed for reconstructing HEP data from granular calorimeters, although currently results are only presented for the MNIST superpixel dataset~\cite{monti2016geometric}.

Another aspect to consider is whether the loss function construction preserves the symmetries of GNN algorithm when predicting unordered sets.
For instance, traditional loss functions like the mean-squared error (MSE) are not invariant with respect to permutations of the output and target sets because the outputs must be reconstructed in the same order as the targets to achieve a small value of the loss function.
To preserve this property, alternative permutation-invariant loss functions like the Chamfer distance~\cite{10.5555/1622943.1622971,Fan_2017_CVPR,Zhang2020FSPool}, Hungarian loss~\cite{zhang2020deepset}, and differential approximations of the Earth mover's distance~\cite{192468,Fan_2017_CVPR,Komiske:2019fks} have been proposed.

\subsection{Computational Performance}
\label{sec:compperf}

One of the most crucial factors in determining the computational performance of a GNN is the graph connectivity. 
The number of edges in a graph usually defines the memory and speed bottleneck, because there are typically more edges than nodes and the $\phi^e$ function is applied the most times. 
If the graph is densely connected, the number of edges is scales quadratically with the number of nodes $N^e \propto (N^v)^2$.
Even without such as severe scaling, if the $\phi^e$ is a large neural network or if there a multiple stacked blocks, the computational resources needed can still be large.
For instance, the tracking GNN of Ref.~\cite{Ju:2020xty} takes as input a portion of a collision event containing approximately 2,500 nodes and 25,000 edges. 
Given the size of the networks and the multiple repeated iterations, one inference requires 52~GFLOPs.
As such, it is imperative to study effective pruning and network compression techniques~\cite{braindamage,deepcompression,energyaware,learningeff,l0reg,lotteryticket}, reduced precision~\cite{Duarte:2018ite,bnnpaper,Coelho:2020zfu}, and alternative hybrid network architectures~\cite{squeezenet,mobilenets,PVCNN} designed to be more efficient.

Another consideration for building and efficiently training GNNs on hardware is whether to use dense or sparse implementations of the graph's adjacency matrix. 
A dense adjacency matrix supports fast, parallel matrix multiplication to compute $E'$, which, for example, is exploited in GCNs and transformers.
However, the adjacency matrix's memory footprint is quadratic in the number of nodes: 10,000 fully-connected nodes corresponds to an adjacency matrix with 100,000,000 entries and thus 400~MB for a 32-bit representation or 12.5~MB with a binary representation. 
Alternatively, using sparse adjacency matrices implies the memory scales linearly in the number of edges, which allows much larger graphs to be processed. 
However, the sparse indexing operations required to implement sparse matrix multiplication can incur greater computational costs than their dense counterparts. 
Such sparse operations are a bottleneck in current deep learning hardware, and next-generation hardware may substantially improve their speed, this would potentially improve the relative advantage of sparse edge implementations of GNNs.

An important advantage of GNN-based approaches over traditional methods for HEP reconstruction is the ability to natively run on highly parallel computing architectures.
All of the deep learning software frameworks for graphs, like PyTorch Geometric~\cite{PyTorchGeometric}, Deep Graph Library~\cite{wang2019dgl}, DeepMind's graph\_nets~\cite{graph_nets} and jraph~\cite{jraph} libraries, StellarGraph~\cite{StellarGraph}, and Spektral~\cite{grattarola2020graph,spektral}, support GPUs to parallelize the algorithm execution.
Work has also been done to accelerate the inference of deep neural networks with field-programmable gate arrays (FPGAs)~\cite{FINN,FINNR,fpgadeep,fpgaover,Duarte:2018ite,Summers:2020xiy,bnnpaper,Coelho:2020zfu}, including GNNs~\cite{Iiyama:2020wap,Heintz:2020soy}, and using heterogeneous computing resources as a service~\cite{Duarte:2019fta,Krupa:2020bwg,Rankin:2020usv}.
Graph processing on FPGAs, reviewed in Ref.~\cite{graphsfpgas}, is a potentially promising direction.
However, we note that detailed and fair comparisons of the computational and physics performance between GNN-based algorithms and traditional HEP algorithms have not yet been extensively performed.
This is a major deliverable of future work.

\section{Applications to Particle Physics Tasks}
\label{sec:gnnapps}

In this section, we review applications of graph neural networks to a variety of reconstruction tasks in high energy physics.
The main graph learning objectives used in HEP reconstruction tasks are 
\begin{itemize}
\item \textit{edge classification}: the prediction of edge-level outputs used to classify edges,
\item \textit{node classification or regression}: the prediction of node-level outputs, representing class probabilities or node properties,
\item \textit{graph pooling}: associating related nodes and edges and possibly predicting properties of these neighborhoods, and
\item \textit{global graph classification}: prediction of a single vector of probabilities the entire graph; this is common for jet and event identification at the LHC and neutrino event classification, but not covered here.
\end{itemize}


\subsection{Charged Particle Tracking}

\label{sec:tracking}

In HEP data analysis, it is crucial to estimate the kinematics of the particles produced in a collision event, such as the position, direction, and momentum of the particles at their production points, as accurately as possible. 
For this purpose, a set of tracking devices (or \emph{trackers}) providing high-precision position measurements is placed close to the beam collision area. 
Charged particles created in the collisions ionize the material of these devices as they exit the collision area, providing several position measurements along the trajectory of each particle. 
To prevent the detector elements from disturbing the trajectory of the particles, the amount of material present in such tracking detectors is kept to a minimum.
The tracker is usually immersed in a strong magnetic field that bends the trajectory, as a means to measure the components of the momentum---the curvature is proportional to the momentum component transverse to the magnetic field.

\begin{figure}
    \centering
    \includegraphics[width=0.4\textwidth]{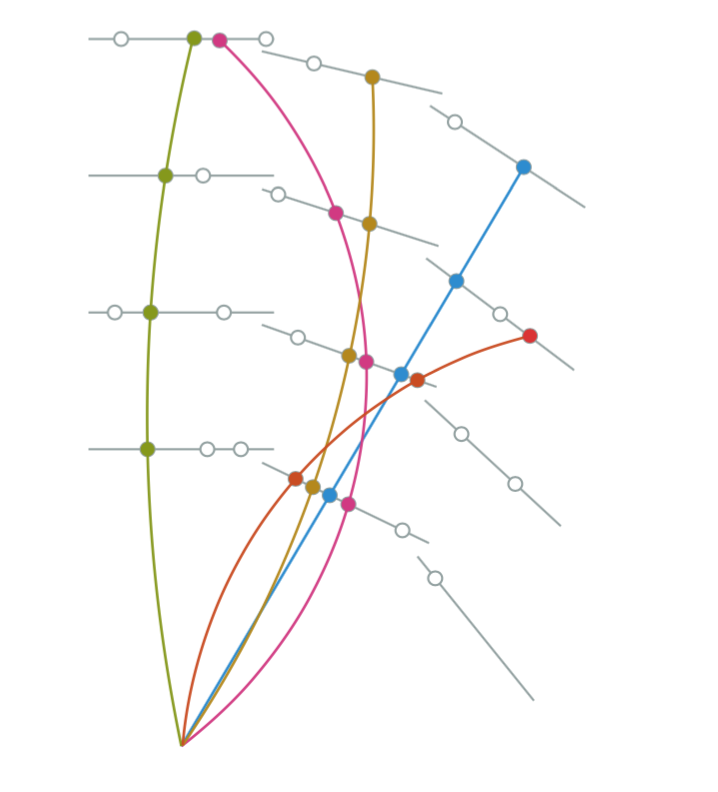}
    \caption{Illustration of the charged particle tracking task~\cite{Amrouche:2019wmx}.
    Each colored curve is the trajectory of a charged particle in a constant magnetic field perpendicular to the viewing plane. 
    The solid circles are hits left by the particle as it traverses the sensitive layers.
    Empty circles are spurious noise hits not created by a reconstructible particle.}
    \label{fig:simpletracking}
\end{figure}

The task of track reconstruction is traditionally divided into two subtasks, track finding and track fitting, although modern techniques may combine them~\cite{Amrouche:2019wmx,Strandlie:2010zz}.
Track finding is a pattern recognition or classification problem and aims at dividing the set of measurements in a tracking detector into subsets (or track candidates) containing measurements believed to originate from the same particle. 
An illustration of a simple track finding problem is shown in Fig.~\ref{fig:simpletracking}. 
It is the task of track finding to associate hits to their respective tracks. 

The track fit takes the set of measurements in a track candidate and estimates as accurately as possible a set of parameters describing the state of the particle somewhere in the tracking detector, often at a reference surface close to the particle beam.
The fitted parameters of the track, especially the curvature, allow for the measurement of the momentum and charge of the particle.
Ideally, each particle would leave one and only one hit on each layer of the detector, the trajectories would be exact helices, and the coordinates would be exact.
In reality, particles may leave multiple hits or no hits in a layer, inhomogeneities in the magnetic field result in distorted arcs, particles may undergo multiple scattering, and the measurements may have anisotropic uncertainties.
Given that these complications are commonplace, a solution that is robust to them is desirable.

Current tracking algorithms include the combinatorial track finder (CTF)~\cite{Chatrchyan:2014fea,Aaboud:2017all} based on the combinatorial Kalman filter~\cite{combkalman1,combkalman2,combkalman3,kalman} that allows pattern recognition and track fitting to occur in the same framework.  
Another tracking algorithm uses a Hough transform~\cite{Aggleton:2017ljc} to identify groups of hits that are roughly consistent with a track hypothesis, reducing
the combinatorial background in the downstream steps.
This algorithm is optimized for the real-time trigger system.
One major computational bottleneck common to many of these algorithms is the combinatorial explosion of possible track candidates, or \emph{seeds}, in high hit density environments.
Improved track seeding, based on global pattern recognition, can dramatically improve the computational performance~\cite{Dietrich:2019qif}.

Lately, there has been increased interest in exploring new methods to address the trade-off between algorithmic quality (good track reconstruction) and speed, which motivated the TrackML particle tracking challenge~\cite{Amrouche:2019wmx,AIHEP:8}
.
From the ML point of view, the problem can be treated as a latent variable problem similar to clustering, in which particle trajectory ``memberships'' must be
inferred, a sequence prediction problem (considering trajectories as time series), a pattern denoising problem treating the sampled trajectories as noisy versions of ideal, continuous traces, or an edge classification problem on graph-encoded hit data. 

\begin{figure}[htbp]
    \centering
    \includegraphics[width=\textwidth]{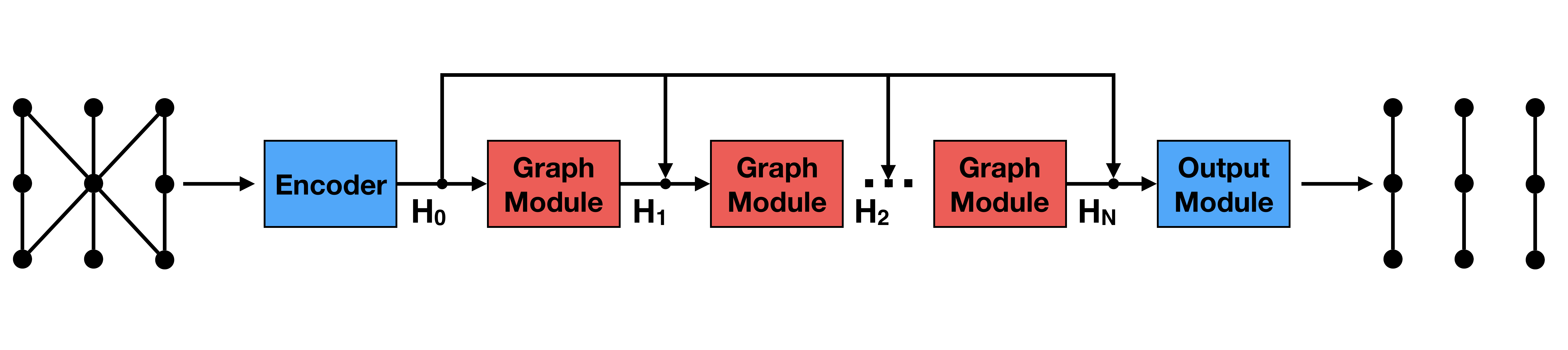}
    \caption{Graph neural network architecture for particle tracking~\cite{Ju:2020xty}. 
    The initial latent features of the nodes and edges after the encoder network are named $H_0$. 
    The graph module is applied repeatedly to the latent features. 
    For the $i$th iteration, the initial features $H_0$ are concatenated with the current features $H_{i-1}$. 
    After $8$ iterations, the output network takes the last latent features $H_8$ to produce classification scores for every edge. 
}
    \label{fig:gnnarch}
\end{figure}

\begin{figure}[htbp]
    \centering
    \includegraphics[width=\textwidth]{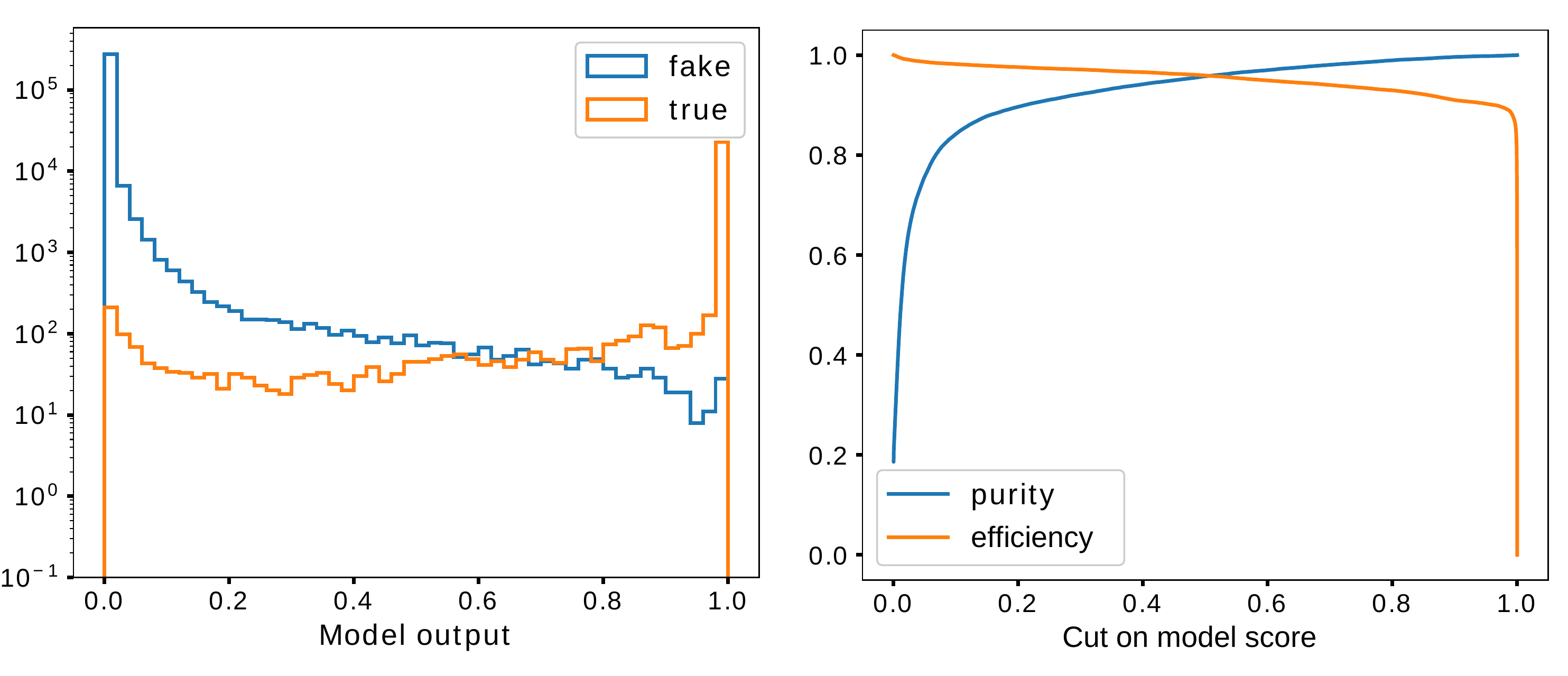}
    \caption{The distribution of the segment classifier scores predicted by the GNN from Ref.~\cite{Ju:2020xty} for true segments (orange) and fake segments (blue), showing clear separation between the two (left). 
    The track segment purity (blue) and efficiency (orange) as a function of different cuts on the model score (right). 
    With a threshold of 0.5 on the GNN output, the edge efficiency, defined as the ratio of the number of true edges passing the threshold over the number of total true edges, reaches 95.9\%, and the purity, defined as the ratio of the number of true edges passing the threshold over the number of total edges passing the threshold, is 95.7\%. }
    \label{fig:tracking_roc}
\end{figure}

The authors of Ref.~\cite{Farrell:2018cjr} propose a GNN approach to charged particle tracking using edge classification.
Each node of the graph represents one hit with edges constructed between pairs of hits on adjacent tracker layers that may plausibly belong to the same track.
After multiple updates of the node representation and edge weights and using the learned edge weight as an attention mechanism, the ``segment classifier'' model learns which edges truly connect hits belonging to the same track.
This approach transforms the clustering problem into an edge classification by targeting the subgraphs of hits belonging to the same trajectories.
This method has high accuracy when applied to a simplified scenario, and is promising for more realistic ones.
In Ref.~\cite{Ju:2020xty} from the same authors, an updated GNN model, based on stacked, repeated interaction network~\cite{Battaglia:2016jem} layers, is presented and provides improved performance.
Figure~\ref{fig:gnnarch} shows the updated architecture, in which the same interaction network layer operates on the initial latent features $H_0$ concatenated with the current features $H_{i-1}$.
After $8$ iterations, the output FC network takes the last latent features $H_8$ to produce classification scores for every edge.
Figure~\ref{fig:tracking_roc} shows the performance of the GNN in correctly classifying the edges, which reaches 95.9\% efficiency and 95.7\% purity on the simulated TrackML dataset~\cite{Amrouche:2019wmx} consisting of top quark-antiquark pairs produced with an additional 200 pileup interactions overlaid to simulate the expected conditions at the HL-LHC.

\subsection{Secondary Vertex Reconstruction}
\label{sec:sv}

The particles that constitute a jet often originate from various intermediate particles that are important to identify in order to fully characterize the jet.
The decay point of the intermediate particle can be identified as a \emph{secondary vertex} (SV), using clustering algorithms on the reconstructed tracks, such as adaptive vertex reconstruction~\cite{Waltenberger:1166320,Fruhwirth:2007hz,5734880}, the CMS inclusive vertex finder~\cite{Sirunyan:2017ezt}, or the ATLAS SV finder~\cite{Heer:2017kbn}.
A review of classical and adaptive algorithms for vertex reconstruction can be found in Ref.~\cite{Strandlie:2010zz}.

Based on the association to a SV, the particles within a jet can be partitioned. 
Properties of the secondary vertices, such as flight distance and total associated energy and mass may then be used in downstream algorithms to identify jets from the decay of bottom or charm quarks.

Through the lens of GNNs, SV reconstruction can be recast as a edge classification and graph partitioning problem.
In Ref.~\cite{Serviansky:2020qwa}, the authors develop a general formalism for set-to-graph (Set2Graph) deep learning and provide mathematical proof that their model formulation is a universal approximator of set-to-graph functions.
In particular, they apply a set-to-edge approximation to the problem of SV reconstruction (particle association) within a jet.
The target is to classify each edge based on whether the two associated particles originate from the same vertex.
The model composes an embedding, a fixed broadcasting map, and a graph-to-graph model to produce the final edge scores.
Though built from simple components, the model's expressivity stems from the equivariant formulation.
Their model outperforms other ML methods, including a GNN~\cite{morris2018weisfeiler}, a Siamese network~\cite{zagoruyko2015learning,NIPS2005_2795,schroff2015facenet}, and a simple multilayer perceptron, on the jet partitioning task by about 10\% in multiple metrics.

\begin{figure}[htbp!]
    \centering
    \includegraphics[width=0.9\textwidth]{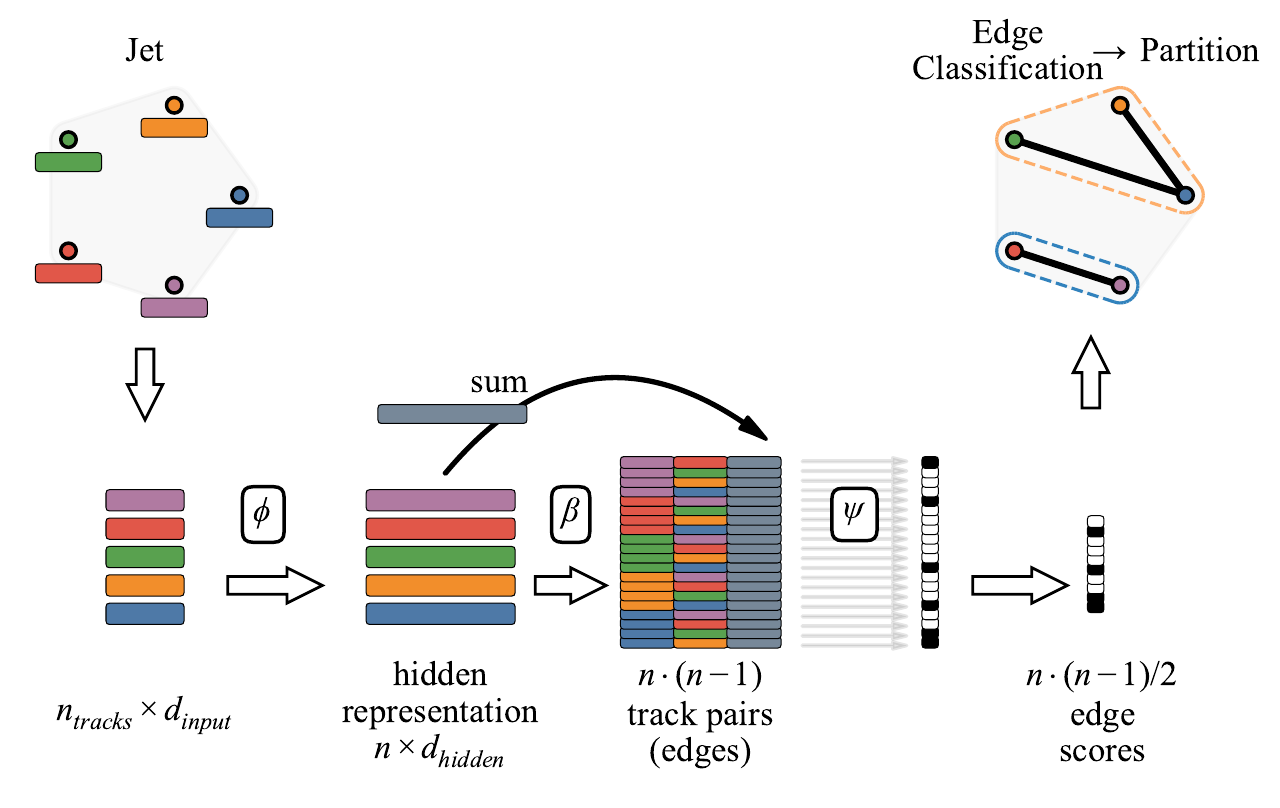}
    \caption{The Set2Graph~\cite{Serviansky:2020qwa,Shlomi:2020ufi} model architecture (top) consisting of a set-to-set component $\phi$, which creates a hidden representation of each track,  a broadcasting layer $\beta$, which creates a representation for each directed edge (ordered pair of tracks in the jet), and an edge classifier $\psi$. 
    Edges whose symmetrized edge score is over a certain threshold are connected, resulting in the set partition. 
    }
    \label{fig:set2graph}
\end{figure}

Ref.~\cite{Shlomi:2020ufi} extends this work and demonstrates the SV reconstruction performance for bottom, charm, and light quark jets, separately, in simulated top quark-antiquark pair events. 
In almost all cases, the Set2Graph model outperforms the standard adaptive vertex reconstruction (AVR) algorithm~\cite{Strandlie:2010zz,Piacquadio:2008zza}, and a simpler, less expressive Set2Graph model called the track pair (TP) classifier.
Figure~\ref{fig:set2graph} shows the Set2Graph model architecture.
The performance may be quantified in terms of the adjusted Rand index (ARI)~\cite{ARI}, which measures the fraction of correctly assigned edges normalized to the expected fraction from random clustering.
They observe a large improvement (33--100\%) in mean ARI for bottom and charm quark jets, and a slight improvement (1\%) for light jets over the AVR and TP classifiers. 

\subsection{Pileup Mitigation}
\label{sec:pileup}

To increase the likelihood of producing rare processes and exotic events, the transverse size of the colliding beams can be squeezed, resulting in multiple interactions per beam crossing. 
The downside of this increased probability is that, when an interesting interaction occurs, it is accompanied by simultaneous spurious interactions (called \textit{pileup}), considered as noise for the analysis.
For instance, the rate of simultaneous interactions per bunch crossing is projected to reach an average of 140--200 for the high-luminosity LHC and 1000 for the proposed 100~TeV hadronic Future Circular Collider (FCC-hh)~\cite{Benedikt:2018csr}.
Pileup increases the likelihood of error in the reconstruction of events of interest because of the contamination from particles produced in different pileup interactions. 
Mitigation of pileup is of prime importance to maintain good efficiency and resolution for the physics objects originating from the primary interaction.
While it is straightforward to suppress charged particles from pileup by identifying their origin, neutral particles are more difficult to suppress.
One of the current state of the art methods is to compute a pileup probability weight per particle~\cite{Bertolini:2014bba} using the local distribution shape, and to use it when computing higher-level quantities.
As a graph-based task, this can generally be conceptualized as a node classification problem.

\begin{figure}
    \centering
    \includegraphics[width=0.8\textwidth]{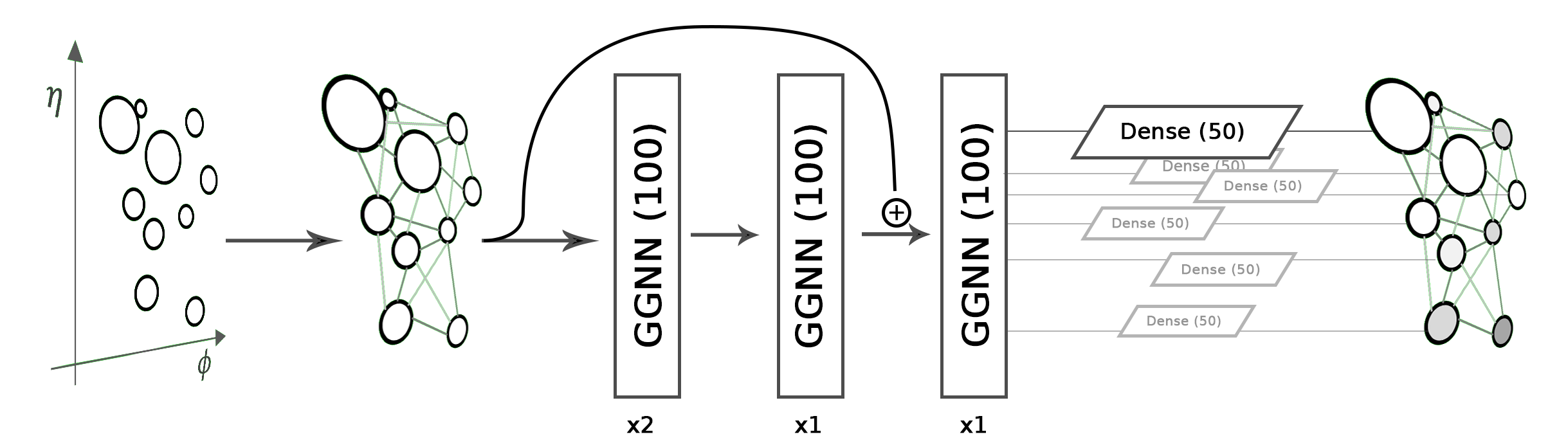}
    \caption{Gated graph network architecture used for pileup mitigation in Ref.~\cite{Martinez:2018fwc}.
    The event is pre-processed by linking local particles together, after which it is fed to 3 gated GNN layers with time steps 2, 1, and 1, respectively, including a residual connection from the first to the third layer. 
    Then a FC network calculates a pileup classification score individually for each graph node.}
    \label{fig:GGNNpileup}
\end{figure}

In Ref.~\cite{Martinez:2018fwc}, the authors utilize the gated GNN architecture~\cite{li2015gated}, shown in Fig.~\ref{fig:GGNNpileup}, to predict a per particle probability of originating from the pileup interactions.
The graph comprises one node per charged and neutral particle of the event, and the edge connectivity is restricted geometrically to $\Delta R < 0.3$ in the $\eta$-$\phi$ plane.
The per-particle pileup probability is extracted with a FC model after three stacked graph layers and a skip connection into the last graph layer.
The model outperforms other methods for pileup subtraction, including GRU and FC network architectures, and improves the resolution of several physical observable.

\begin{figure}
    \centering
    \includegraphics[width=0.9\textwidth]{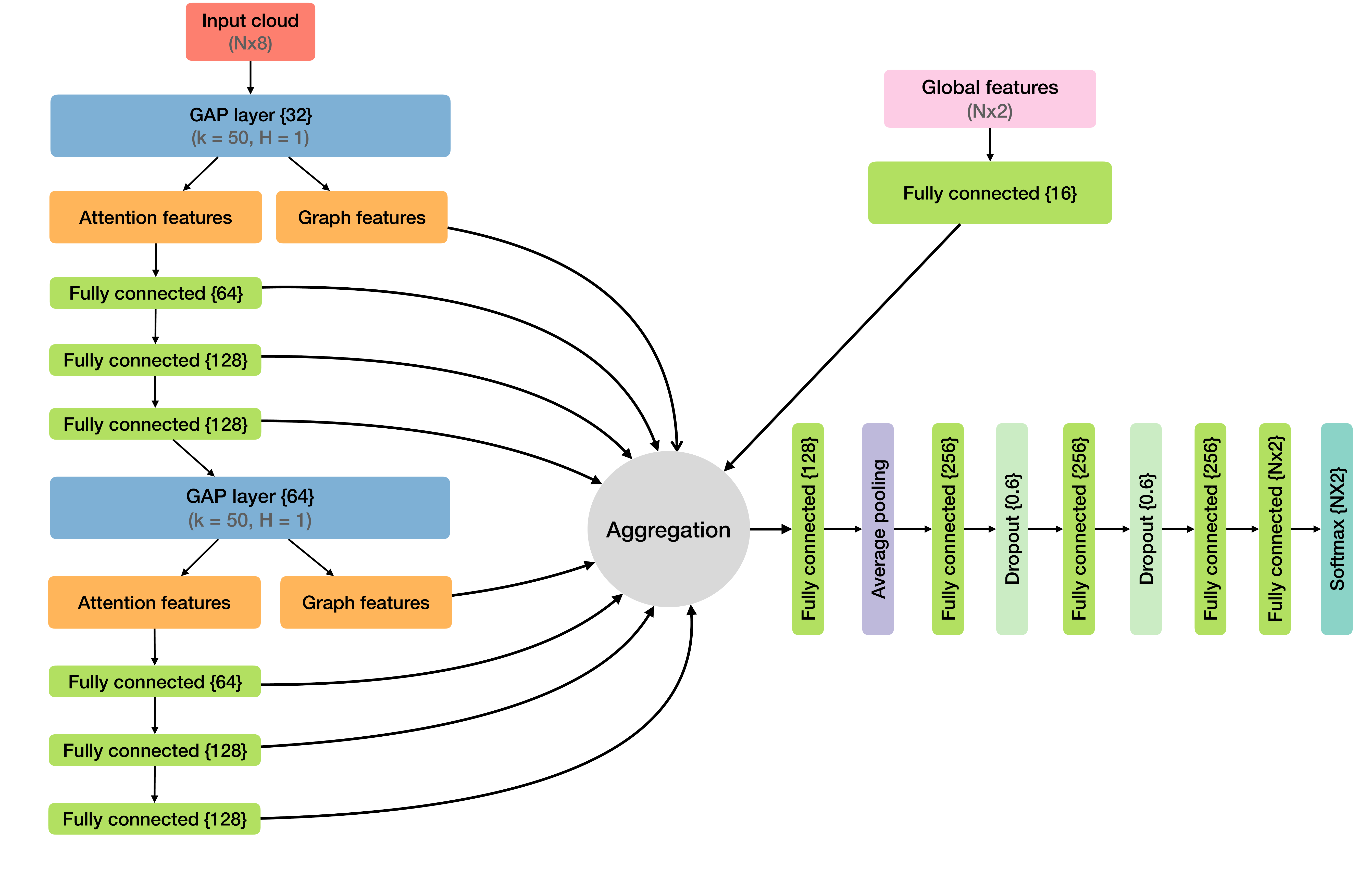}
    \caption{ABCNet architecture used for pileup identification in Ref.~\ref{fig:ABCNet}. 
    FC layer and encoding node sizes are denoted inside curly brackets. 
    For each graph attention pooling layer (GAPLayer), the number of $k$-nearest neighbors and attention heads ($h$) are given.}
    \label{fig:ABCNet}
\end{figure}

The authors of Ref.~\cite{Mikuni:2020wpr} take inspiration from the graph attention network~\cite{gat} and the graph attention pooling network (GAPNet)~\cite{chen2019gapnet} to predict a per-particle pileup probability with a model called attention-based cloud network (ABCNet) shown in Fig.~\ref{fig:ABCNet}.
The node and edge features are updated by multiple FC models, where each (directed) edge is weighted by an attention factor.
The connectivity is initialized to the $k$-nearest neighbors in the feature space then updated based on the latent space of the stacked graph layers.
A multi-head attention mechanism, described in Sec.~\ref{sec:gnns}, is used to improve the robustness of models.
Skip connections further facilitate the information flow.
A global graph latent representation is used to compute an output for each node using a fixed ordering.
This method improves the resolution of the single jet and dijet mass observables over a large range of number of pileup interactions.

\subsection{Calorimeter Reconstruction}
\label{sec:calo}

A \textit{calorimeter} is a block of instrumented material in which particles to be measured are fully absorbed and their energy transformed into a measurable quantity. 
Typically, the interaction of the incident particle with the detector produces a cascade of secondary particles (known as a \emph{shower}) with progressively smaller energies.
The energy deposited by the showering particles in the calorimeter can be detected in the form of charge or light and serves as a measurement of the energy of the incident particle.
There are two primary categories of particle showers, one caused by the electromagnetic force and consisting of electrons, positrons, and photons, and the other resulting from the strong nuclear force and composed of charged and neutral hadrons. 
Corresponding to these two types of particle showers, there the two primary forms of calorimeters: electromagnetic and hadron calorimeters.

Calorimeters can be further classified into sampling and homogeneous calorimeters.
Sampling calorimeters consist of alternating layers of an \emph{absorber}, a dense material used to induce the shower and energy loss of the incident particle, and an active medium that provides the detectable signal. 
Conversely, homogeneous calorimeters are built of one type of material that performs both tasks, energy degradation and signal generation.
Nonetheless, both types are usually segmented into different cells, providing some spatial resolution.
Moreover, reconstruction of the energy of the incoming particle in a calorimeter requires joint clustering and calibration of the signal in various cells.
Reviews of classical techniques for calorimetry in high energy physics can be found in Ref.~\cite{fabjan:1982,Wigmans:1990ey,Fabjan:2003aq}.
From an GNN perspective, calorimeter reconstruction can be thought of as (possible) graph pooling and node regression.

Ref.~\cite{Qasim:2019otl} proposes a GNN-based approach to cluster and assign signals in a high granularity calorimeter to separate particles.
A latent edge representation is constructed using a potential function of the Euclidean distance $d_{jk}$ between nodes $j$ and $k$ in (a subspace of) the latent space
\begin{align}
V_n(d_{jk}) &= \exp(-|d_{jk}|^n) 
\end{align}
as an attention weight.
One proposed model---GravNet---connects the nearest neighbors in a latent space and uses the potential $V_2$, while another---GarNet---uses a fixed number of additional nodes to define the graph connectivity and $V_1$ as the potential.
Node features are updated using the concatenated messages from multiple aggregations, and the output predicts the fraction of a cell's energy belonging to each particle.
These methods improve over classical approaches and could be more beneficial in future detectors with greater complexity.

\begin{figure}[htpb]
    \centering
    \includegraphics[width=0.64\textwidth]{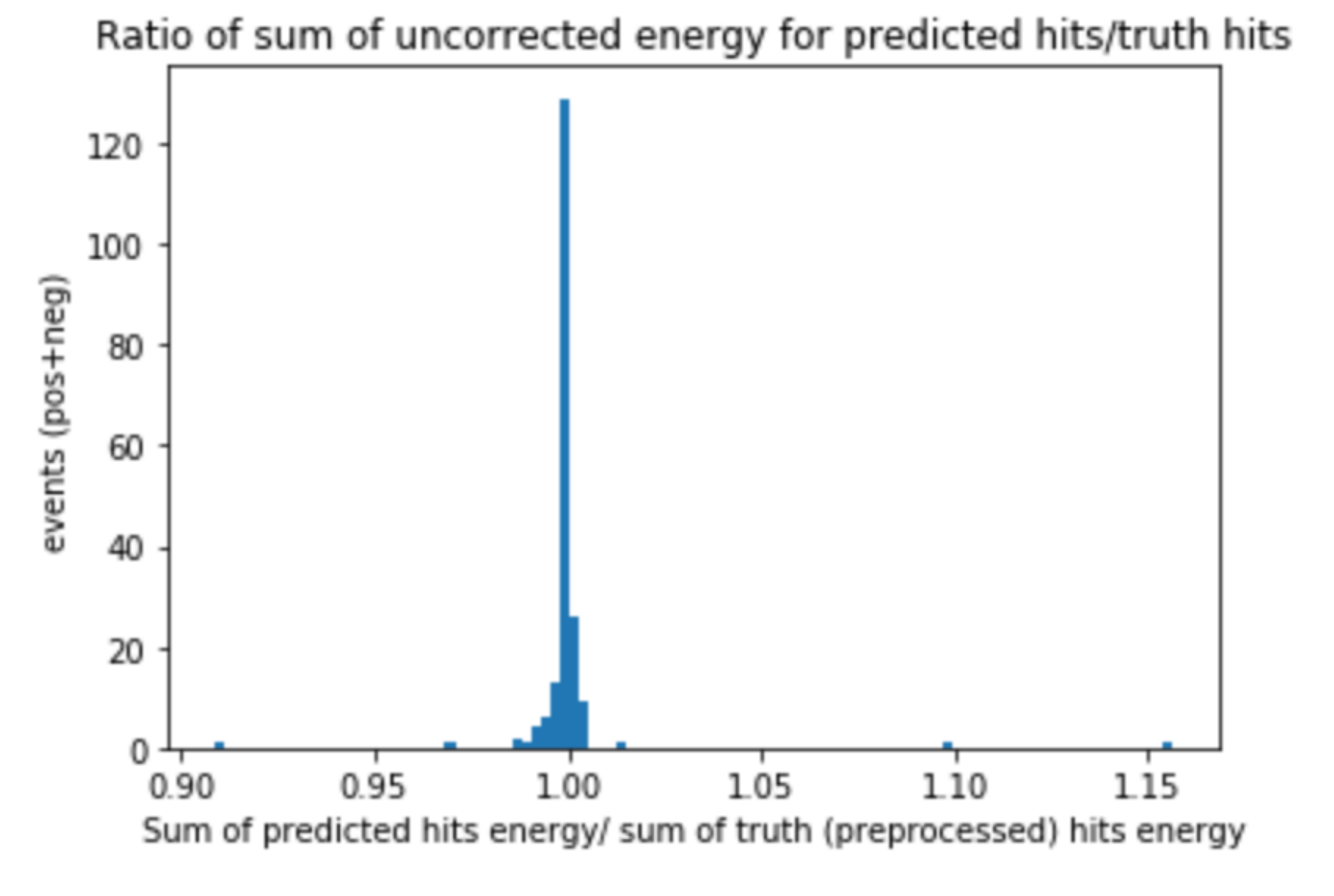}
    \includegraphics[width=0.35\textwidth]{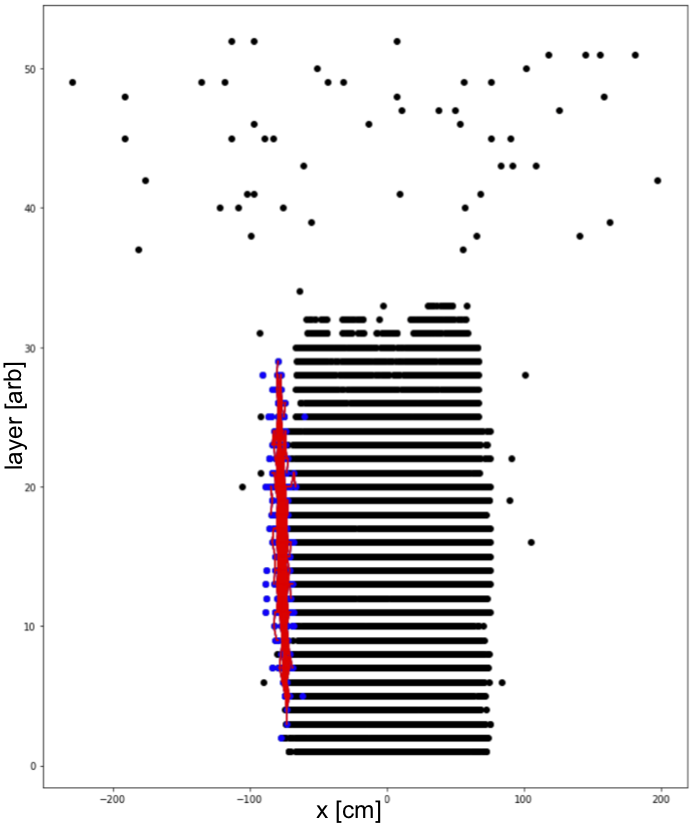}
    \caption{The ratio, per event, for photons of total collected calorimeter energy deposits connected by predicted edges to the energy collected by the associations from ground truth (left) for a GNN from Ref.~\cite{Ju:2020xty}. . 
    The event display of a single photon showing the predicted edges in red, the truth nodes in blue, and the energy deposits from noise in black (right).
    \label{fig:calo_photon}}
\end{figure}

Ref.~\cite{Ju:2020xty} also proposes a GNN approach using stacked EdgeConv layers to identify clusters in the CMS high granularity calorimeter. 
The output is a set of edge weights classifying hit pairs as being particles or noise.
Results are promising in that muons, photons, and pions are efficiently and purely reconstructed and their energy is accurately measured as shown in Fig.~\ref{fig:calo_photon} in the case of photons.
Ongoing work includes studies on how to reconstruct multiple particle types simultaneously using network architectures that can assign categories to edges, and how to deal with overlapping showers and fractional assignment of hit energy into clusters.

\subsection{Particle-Flow Reconstruction}

\label{sec:pf}
\begin{figure}
    \centering
    \includegraphics[width=\textwidth]{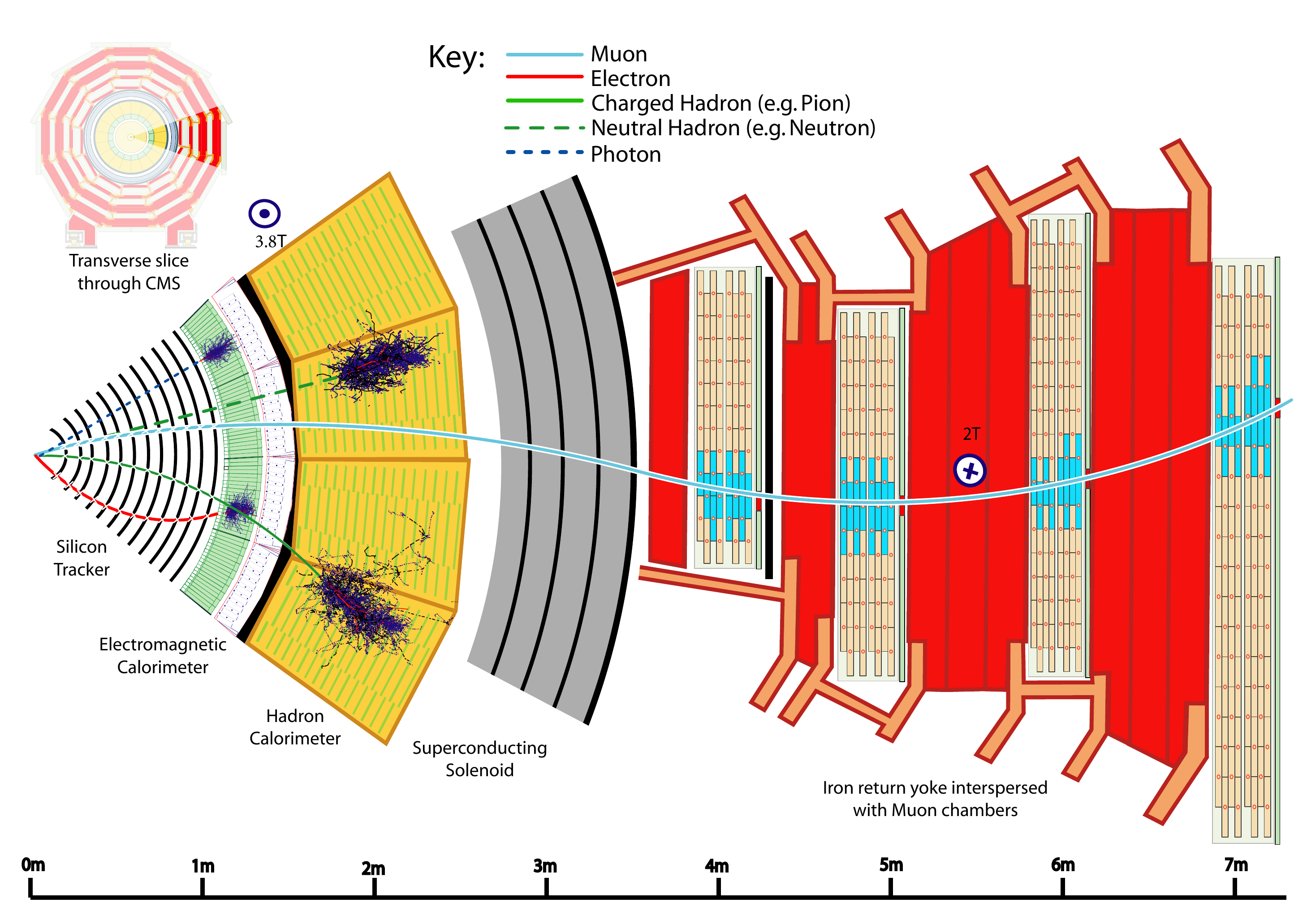}
    \caption{Different types of particles and their corresponding signatures in the CMS detector~\cite{Sirunyan:2017ulk}. 
    Particle-flow algorithms aim to optimally combine different measurements for different subdetectors to reconstruct a list of final-state particles.}
    \label{fig:particleflow}
\end{figure}

Modern general-purpose detectors at high-energy colliders are composed of different types of detector layers nested around the beam axis in addition to forward and backward ``endcap'' layers.
Charged particle tracks are measured by a tracking detector as described in Sec.~\ref{sec:tracking}.
As described in Sec.~\ref{sec:calo}, electrons and photons are absorbed in an electromagnetic calorimeter (ECAL), creating \emph{clusters} of energy that can be measured.
Similarly, charged and neutral hadrons are absorbed, clustered, and measured in a hadron calorimeter (HCAL).
Muons may produce hits in additional tracking layers called muon detectors, located outside of the calorimeters, while neutrinos escape unseen.
Figure~\ref{fig:particleflow} displays a sketch of a transverse slice of a modern general-purpose detector, the CMS detector~\cite{cms_paper} at the CERN Large Hadron Collider (LHC), with different types of particles and their corresponding signatures.

\begin{figure}
    \centering
    \includegraphics[width=\textwidth]{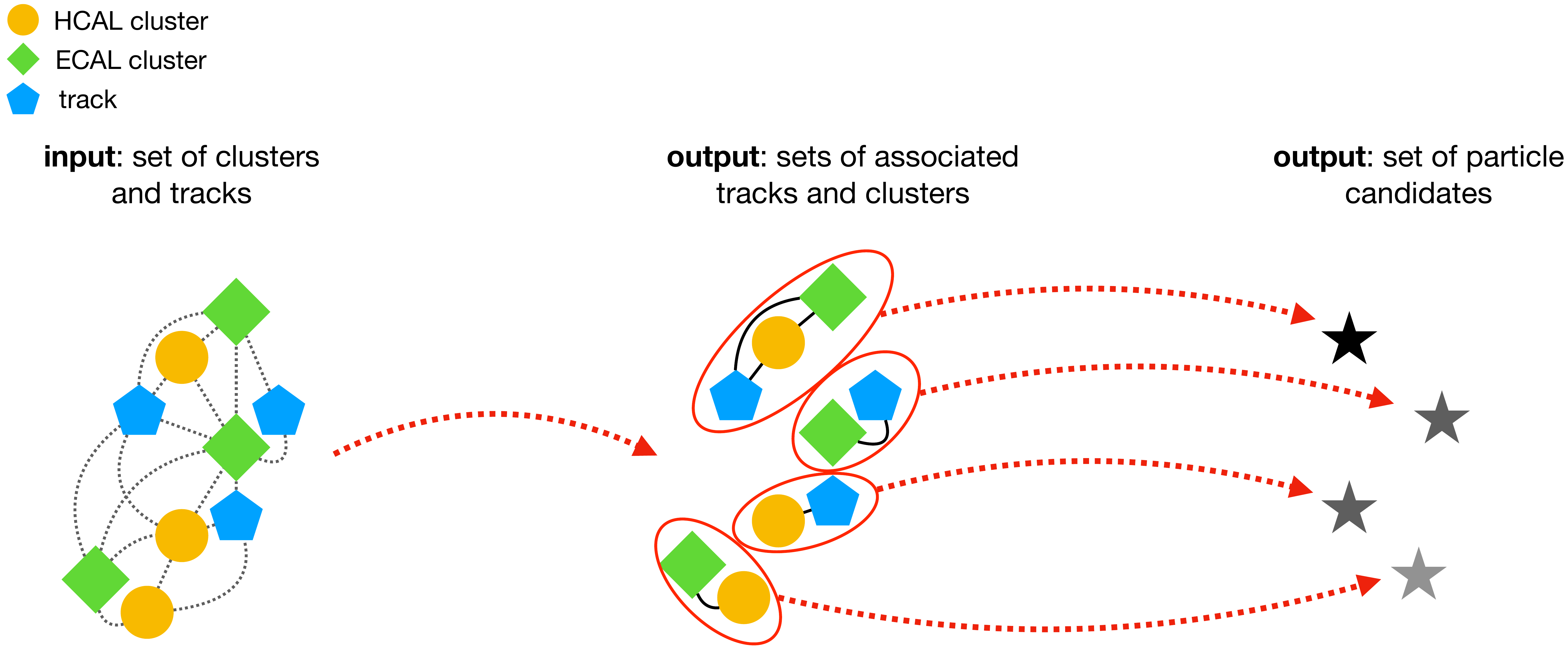}
    \caption{Schematic representation of a particle-flow algorithm based on input HCAL cluster, ECAL clusters, and tracks. 
    These inputs are associated to one another and the list of final-state particles is determined by combining these measurements.}
    \label{fig:PF2}
\end{figure}

An improved global event description can be achieved by correlating the basic \textit{elements} from all detector layers (tracks and clusters) to identify each final-state particle, and by combining the corresponding measurements to reconstruct the particle properties.
This holistic approach is called \textit{particle-flow (PF) reconstruction}.
The PF concept was developed and used for the first time by the ALEPH experiment at LEP~\cite{Buskulic:1994wz} and has been successfully deployed at the LHC in both CMS~\cite{Sirunyan:2017ulk} and ATLAS~\cite{Aaboud:2017aca}.
An important ingredient in this approach is the fine spatial granularity of the detector layers. 
The ultimate goal of PF reconstruction is to provide a complete list of identified final-state particles, with their momenta optimally reconstructed from a combined fit of all pertaining measurements, and links to contributing elements.
From this list of particles, the physics objects can then be determined with superior efficiencies and resolutions.
This is shown schematically in Fig.~\ref{fig:PF2}.

ML methods based on an image representations have been studied for PF reconstruction.
Based on a computer-vision approach, Ref.~\cite{DiBello:2020bas} uses a CNN with up and down sampling via choice of kernel size and stride to combine information from ECAL and HCAL layers to better reconstructed the energies of hadron showers.
As a graph-based learning problem, PF reconstruction has multiple objectives: graph pooling or edge classification for associating input measurements to output particles and node regression for measuring particle momenta.

\begin{figure}[hbtp]
    \centering
    \includegraphics[width=0.49\textwidth]{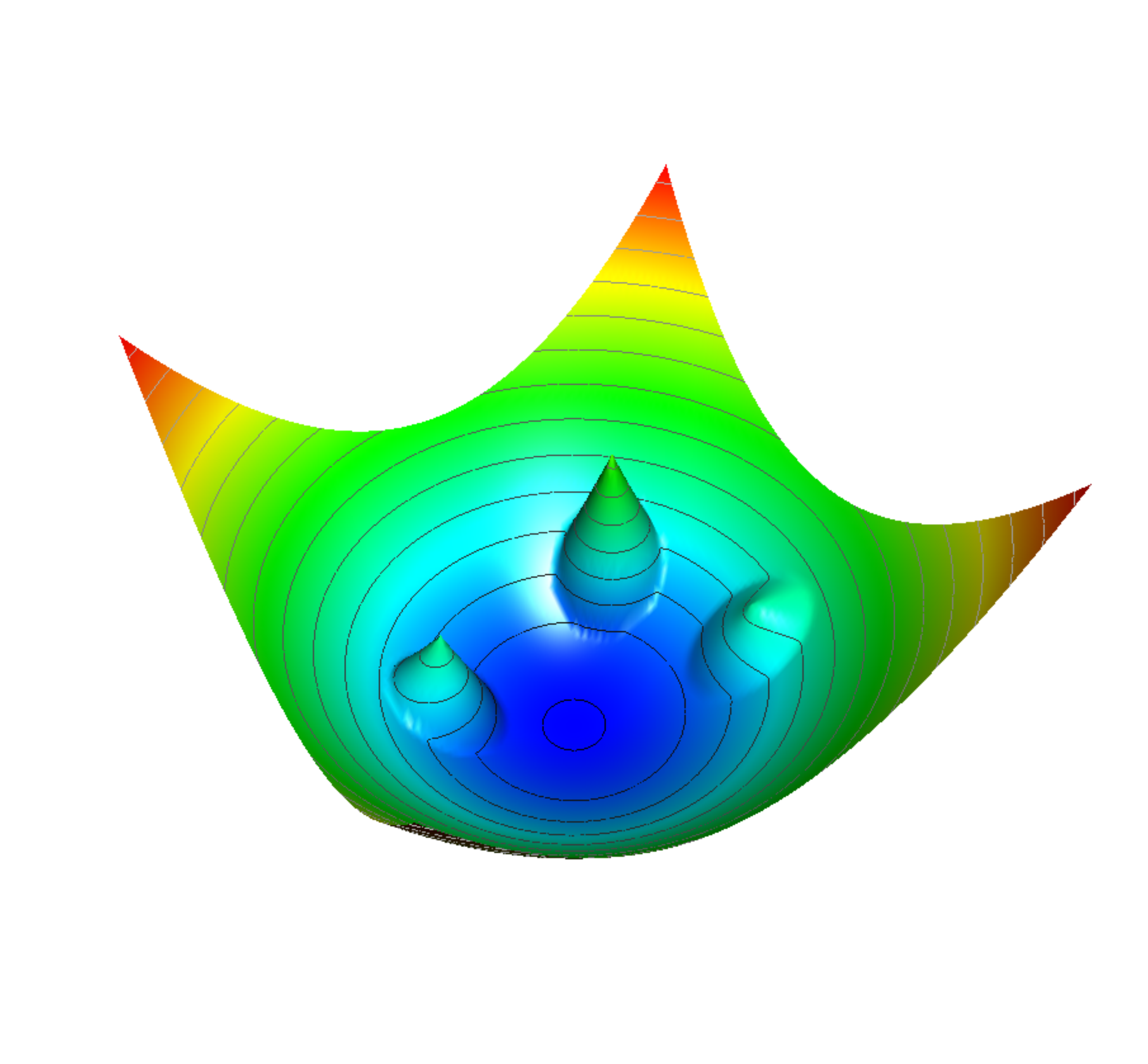}
    \caption{Illustration of the object condensation loss function combining four effective potentials: three that repel a given node and one in the center that attracts the node~\cite{Kieseler:2020wcq}.}
    \label{fig:potential}
\end{figure}

Ref.~\cite{Kieseler:2020wcq} proposes the \textit{object condensation} loss formulation using GNN methods to extract the particle information from the graph of measurements as well as grouping of the measurements.
The model predicts the properties of a smaller number of particles than there are measurements, in essence reducing the graph without explicit assumptions on the number of targeted particles.
Certain nodes are chosen to be the ``condensation'' point of a particle, to which the target properties are attached.
A stacked GravNet model performs node-level regression of a kinematic correction factor together with a \textit{condensation weight} $\beta_i$, whcih indicates whether that node is representative of a particle in the event.
A special loss function mimics attractive and repulsive electromagnetic potentials to ensure nodes belonging to the same particle are close in the latent space.
Explicitly, an effective charge is computed from the condensation weight through a function with zero gradient at 0 and monotonically increasing gradient towards a pole at 1: $q_i = \arctanh^2{\beta_i} + q_\mathrm{min}$. 
The node $\alpha$ with maximum charge $q_\alpha$ for each particle is used to define an attractive potential $\breve{V}_k(x) = ||x-x_\alpha||^2 q_{\alpha k}$ or a repulsive potential $\hat{V}_k(x) = \max(0, 1-||x-x_\alpha||) q_{\alpha k}$ depending on if the node $\alpha$ belongs to the same particle.
This is combined in the loss function,
\begin{align}
\label{eq:potential_loss}
    L_V &= \frac{1}{N}\sum_{j=1}^N q_j \sum_{k=1}^K \left( M_{jk}\breve{V}_k(x_j) + (1-M_{jk})\hat{V}_k(x_j)  \right) ,
\end{align}
where $M_{jk}$ is 1 if node $j$ belongs to particle $k$ and 0 otherwise.
As illustrated in Fig.~\ref{fig:potential}, apart from a few saddle points, the node is pulled towards the nodes belonging to the same particle and away from nodes belonging to other particles.

The performance of this algorithm is compared with a baseline PF algorithm in a sparse, low-pileup LHC environment.
The proposed method selects more real particles and misidentifies less fake particles than the standard approach.

\section{Summary}
\label{sec:summary}

Graph neural networks (GNNs) that operate on point clouds and graphs are increasingly popular for applications in high energy physics (HEP) event reconstruction.
One reason for their popularity is a closer correspondence to the input HEP data or the desired output.
Namely, measurements in a detector naturally form a point cloud, which can be interpreted as the nodes in a graph once the connectivity (edges) is specified.
The solution to many HEP reconstruction tasks can be mapped onto the edges of the graph (e.g. track finding), the nodes of the graph (e.g. pileup mitigation), or graph characteristics (e.g. jet tagging).
Another reason is practical: the computational performance of many traditional reconstruction approaches scales poorly as the collision events become more complex, while GNNs have the potential to scale up better, especially by leveraging highly parallel architectures like graphics processing units or field-programmable gate arrays.

A variety of GNN models have been used for node-level, edge-level, and graph-pooled tasks, and all models share common structures that involve propagating and aggregating information between different nodes in the graph.
Another key ingredient is in the construction of the initial graph connectivity and whether that connectivity is dynamic (learned) or static. 
The physics performance of GNNs has been shown to match or surpass that of state-of-the-art techniques in several proof-of-concept studies.
However, many of the models have not yet been tested with real detector data, or benchmarked in terms of their computational performance.
Nonetheless, the approach is increasingly promising, as more and more HEP applications continue to appear.
At their core, GNNs model the nature of the interactions between the objects in an input set, which may explain why particle physicists, trying to model the nature of the interactions between elementary particles, find them so applicable.

\section*{Acknowledgments}
We thank Jonathan Shlomi and Peter Battaglia for discussions and sharing materials reproduced here.
We thank authors of other chapters for feedback on this one.
J.~D. is supported by the U.S. Department of Energy (DOE), Office of Science, Office of High Energy Physics Early Career Research program under Award No. DE-SC0021187.
J-R.~V. is partially supported by the European Research Council (ERC) under the European Union's Horizon 2020 research and innovation program (Grant Agreement No. 772369) and by the U.S. DOE, Office of Science, Office of High Energy Physics under Award No. DE-SC0011925, DE-SC0019227, and DE-AC02-07CH11359.

\bibliographystyle{tepml}
\bibliography{ws-rv-sample}


\end{document}